\documentclass{article}
\usepackage{microtype}
\usepackage{graphicx}
\usepackage{booktabs} 

\usepackage[accepted]{icml2018} 

\usepackage[utf8]{inputenc} 
\usepackage[T1]{fontenc}    
\usepackage{hyperref}       
\usepackage{url}            
\usepackage{booktabs}       
\usepackage{amsfonts}       
\usepackage{nicefrac}       
\usepackage{microtype}      
\usepackage{subcaption}
\usepackage{helvet}
\usepackage{courier}
\usepackage{multirow}
\usepackage{xcolor}
\usepackage{url}
\usepackage{amsmath,amscd,amsbsy,amssymb,latexsym,url,bm}
\usepackage{graphicx}
\usepackage{enumitem,balance}
\usepackage{wrapfig}
\usepackage{mathrsfs, euscript}
\usepackage[colorinlistoftodos]{todonotes}
\usepackage{ifpdf}
\usepackage{makecell}
\usepackage{tikz}
\usetikzlibrary{matrix,chains,positioning,decorations.pathreplacing,arrows}
\usepackage{amsthm}
\usepackage{footmisc}
\usepackage[english]{babel}
\usepackage[utf8]{inputenc}
\usepackage{thmtools,thm-restate}
\newtheorem{theorem}{Theorem}

\newtheorem{lemma}{Lemma}
\newtheorem{assumption}{Assumption}
\newtheorem{property}{Property}

\usepackage[outercaption]{sidecap}    

\usepackage{enumitem}
\usepackage{ragged2e}

\usepackage{mathtools}

\usepackage{amssymb}
\usepackage[cal=boondoxo,calscaled=1]{mathalfa}
\usepackage[rmdefault=ntxmi,sfdefault=iwona,scaled=1.0]{isomath}
\DeclareSymbolFont{letters}{OML}{ntxmi}{m}{it}
\DeclareMathOperator{\diag}{diag}

\newcommand*\vs[1]{\vectorsym{#1}}

\DeclareRobustCommand{\frac}[3][0pt]{{\begingroup\hspace{#1}#2\hspace{#1}\endgroup\over\hspace{#1}#3\hspace{#1}}}
\newcommand*\nash{\mathop{}\!\mathtt{Nash}}
\newcommand*\mf{\mathop{}\!\mathtt{MF}}

\begin{document}

\twocolumn[
\icmltitle{Mean Field Multi-Agent Reinforcement Learning}

\icmlsetsymbol{equal}{*}

\begin{icmlauthorlist}
	\icmlauthor{Yaodong Yang}{ucl}
	\icmlauthor{Rui Luo}{ucl}
	\icmlauthor{Minne Li}{ucl}
	\icmlauthor{Ming Zhou}{sjtu}
	\icmlauthor{Weinan Zhang}{sjtu}
	\icmlauthor{Jun Wang}{ucl}
\end{icmlauthorlist}

\icmlaffiliation{ucl}{University College London, London, U.K.}
\icmlaffiliation{sjtu}{Shanghai Jiao Tong University, Shanghai, China}

\icmlcorrespondingauthor{Yaodong Yang}{yaodong.yang@cs.ucl.ac.uk} 
\icmlcorrespondingauthor{Jun Wang}{j.wang@cs.ucl.ac.uk} 

\icmlkeywords{Mean Field Reinforcement Learning, Multi-Agent Reinforcement Learning}

\vskip 0.3in
]



\printAffiliationsAndNotice{} 


\begin{abstract}
Existing multi-agent reinforcement learning methods are limited typically to a small number of agents. When the  agent number increases largely, the learning becomes intractable due to the curse of the dimensionality and the exponential growth of agent interactions. In this paper, we present \emph{Mean Field Reinforcement Learning} where the interactions within the population of agents are approximated by those between a single agent and the average effect from the overall population or neighboring agents; the interplay between the two entities is mutually reinforced: the learning of the individual agent's optimal policy depends on the dynamics of the population, while the dynamics of the population change according to the collective patterns of the individual policies. We develop practical mean field Q-learning and mean field Actor-Critic algorithms and analyze the convergence of the solution to Nash equilibrium. Experiments on Gaussian squeeze, Ising model, and battle games justify the learning effectiveness of our mean field approaches. In addition, we report the first result to solve the Ising model via model-free reinforcement learning methods.
\end{abstract}


\section{Introduction}\label{sec:intro}

Multi-agent reinforcement learning (MARL) is concerned with a set of autonomous agents that share a common environment \cite{busoniu2008comprehensive}. Learning in MARL is fundamentally difficult since agents not only interact with the environment but also with each other. Independent $Q$-learning \cite{tan1993multi} that considers other agents as a part of the environment often fails as the multi-agent setting breaks the theoretical convergence guarantee and makes the learning unstable: changes in the policy of one agent will affect that of the others, and vice versa \cite{matignon2012independent}. 


Instead, accounting for the extra information from \emph{conjecturing} the policies of other agents is beneficial to each single learner \cite{DBLP:journals/corr/abs-1709-04326,lowe2017multi}. 
Studies show that an agent who learns the effect of joint actions has better performance than those who do not in many scenarios, including cooperative games \cite{panait2005cooperative}, zero-sum stochastic games \cite{littman1994markov}, and general-sum stochastic  games \cite{littman2001friend, hu2003nash}.

The existing equilibrium-solving approaches, although principled, are only capable of solving a handful of agents \cite{hu2003nash,bowling2002multiagent}. The computational complexity of directly solving (Nash) equilibrium would prevent them from applying to the situations with a large group or even a population of agents. Yet, in practice, many cases do require strategic interactions among a large number of agents, such as the gaming bots in  
Massively Multiplayer Online Role-Playing Game
\cite{jeong2015analysis}, the trading agents in stock markets \cite{troy1997envisioning}, or the online advertising bidding agents \cite{wang2017display}.  

In this paper, we tackle MARL when a large number of agents co-exist. We consider a setting where each agent is directly interacting with a finite set of other agents; through a chain of direct interactions, any pair of agents is interconnected globally \cite{blume1993statistical}. The scalability is solved by employing Mean Field Theory \cite{stanley1971phase} -- the interactions within the population of agents are approximated by that of a single agent played with the average effect from the overall (local) population. The learning is mutually reinforced between two entities rather than many entities: the learning of the individual agent's optimal policy is based on the dynamics of the agent population, meanwhile, the dynamics of the population is updated according to the individual policies. Based on such formulation, we develop practical mean field $Q$-learning and mean field Actor-Critic algorithms, and discuss the convergence of our solution under certain assumptions.  Our experiment on a simple multi-agent resource allocation shows that our mean field MARL is capable of learning over many-agent interactions when others fail. We also demonstrate that with temporal-difference learning, mean field MARL manages to learn and solve the Ising model without even explicitly knowing the energy function. At last, in a mixed cooperative-competitive battle game, we show that the mean field MARL achieves high winning rates against other baselines previously reported for many agent systems. 




\section{Preliminary}

MARL intersects between reinforcement learning and game theory. 
The marriage of the two gives rise to the general framework of \emph{stochastic game} \cite{shapley1953stochastic}.

\subsection{Stochastic Game}

An $N$-agent (or, $N$-player) stochastic game $\Gamma$ is formalized by the tuple $\Gamma \triangleq \big( \mathcal{S}, \mathcal{A}^1, \dots, \mathcal{A}^N, r^1, \dots, r^N, p, \gamma \big)$, where $\mathcal{S}$ denotes the state space, and $\mathcal{A}^j$ is the action space of agent $j \in \{1,\dots,N\}$. The reward function for agent $j$ is defined as $r^j : \mathcal{S}\times \mathcal{A}^1\times \dots \times \mathcal{A}^N \to \mathbb{R}$, determining the immediate reward. The transition probability $p : \mathcal{S}\times \mathcal{A}^1\times \dots \times \mathcal{A}^N \to \Omega(\mathcal{S})$ characterizes the stochastic evolution of states in time, with $\Omega(\mathcal{S})$ being the collection of probability distributions over the state space $\mathcal{S}$. The constant $\gamma\in [0,1)$ represents the reward discount factor across time. At time step $t$, all agents take actions simultaneously, each receives the immediate reward $r^j_t$ as a consequence of taking the previous actions.

The agents choose actions according to their policies, also known as strategies. For agent $j$, the corresponding policy is defined as $\pi^j : \mathcal{S} \to \Omega(\mathcal{A}^j)$, where $\Omega(\mathcal{A}^j)$ is the collection of probability distributions over agent $j$'s action space $\mathcal{A}^j$. 
Let $\vs{\pi} \triangleq [\pi^1, \dots, \pi^N]$ denote the joint policy of all agents; 
we assume, as one usually does, $\vs{\pi}$ to be time-independent, which is referred to be \emph{stationary}.
Provided an initial state $s$, the value function of agent $j$ under the joint policy $\vs{\pi}$ is written as the expected cumulative discounted future reward:
\vskip -0.25in
\begin{align}
\label{eq:v}
v^j_{\vs{\pi}}(s) = v^j(s; {\vs{\pi}}) = \sum_{t=0}^\infty \gamma^t \mathbb{E}_{\vs{\pi}, p} \big[ r^j_t | s_0 = s, \vs{\pi} \big].
\end{align}
\vskip -0.15in
The $Q$-function (or, the action-value function) can then be defined within the framework of $N$-agent game based on the Bellman equation given the value function in Eq.~\eqref{eq:v} such that the $Q$-function $Q^j_{\vs{\pi}}: \mathcal{S}\times \mathcal{A}^1\times \dots \times \mathcal{A}^N \to \mathbb{R}$ of agent $j$ under the joint policy $\vs{\pi}$ can be formulated as
\begin{align}
\label{eq:q}
Q^j_{\vs{\pi}}(s, \vs{a}) &= r^j(s, \vs{a}) + \gamma \mathbb{E}_{s' \sim p} [v^j_{\vs{\pi}}(s')]~,
\end{align}
where $s'$ is the state at the next time step. The value function $v^j_{\vs{\pi}}$ can be expressed in terms of the $Q$-function in Eq.~\eqref{eq:q} as
\begin{align}
\label{eq:v2}
v^j_{\vs{\pi}}(s) = \mathbb{E}_{\vs{a} \sim \vs{\pi}} \big[ Q^j_{\vs{\pi}}(s, \vs{a}) \big].
\end{align}
The $Q$-function for $N$-agent game in Eq.~\eqref{eq:q} extends the formulation for single-agent game by considering the joint action taken by all agents $\vs{a} \triangleq [a^1, \dots, a^N]$, and by taking the expectation over the joint action in Eq.~\eqref{eq:v2}.

We formulate MARL by the stochastic game with a discrete-time non-cooperative setting, \emph{i.e.} no explicit coalitions are considered. The game is assumed to be incomplete but to have perfect information \cite{littman1994markov}, \emph{i.e.} each agent knows neither the game dynamics nor the reward functions of others, but it is able to observe and react to the previous actions and the resulting immediate rewards of other agents.


\subsection{Nash $Q$-learning}


In MARL, the objective of each agent is to learn an optimal policy to maximize its value function. Optimizing the $v^j_{\vs{\pi}}$ for agent $j$ depends on the joint policy $\vs{\pi}$ of all agents, the concept of \emph{Nash equilibrium} in stochastic games is therefore of great importance \cite{hu2003nash}. It is represented by a particular joint policy $\vs{\pi_*} \triangleq [\pi^1_*, \dots, \pi^N_*]$ such that for all $s \in \mathcal{S},~j \in \{1,\dots,N\}$ and all valid $\pi^j$, it satisfies
\begin{align*}
v^j(s; \vs{\pi}_*) = v^j(s ; \pi^j_*, \vs{\pi}^{-j}_*) \ge v^j(s ; \pi^j, \vs{\pi}^{-j}_*).
\end{align*}
Here we adopt a compact notation for the joint policy of all agents except $j$ as $\vs{\pi}^{-j}_* \triangleq [\pi^1_*, \dots, \pi^{j-1}_*, \pi^{j+1}_*, \dots, \pi^N_*]$.

In a Nash equilibrium, each agent acts with the \emph{best response} $\pi^j_*$ to others, provided that all other agents follow the policy $\vs{\pi}^{-j}_*$. It has been shown that, for a $N$-agent stochastic game, there is at least one Nash equilibrium with stationary policies \cite{fink1964equilibrium}. Given a Nash policy $\vs{\pi}_*$, the Nash value function $\vs{v}^{\nash}(s) \triangleq [v^1_{\vs{\pi}_*}(s),\dots,v^N_{\vs{\pi}_*}(s)]$ is calculated with all agents following $\vs{\pi}_*$ from the initial state $s$ onward.
Nash $Q$-learning \cite{hu2003nash} defines an iterative procedure with two alternating steps for computing the Nash policy: 1) solving the Nash equilibrium of the current stage game defined by $\{\vs{Q}_t\}$ using the Lemke-Howson algorithm \cite{lemke1964equilibrium}, 2)
improving the estimation of the $Q$-function with the new Nash equilibrium  value. 
It can be proved that under certain assumptions, the Nash operator $\mathcal{H}^{\nash}$ defined by the following expression  
\begin{align}
\mathcal{H}^{\nash} \vs{Q}(s, \vs{a}) = \mathbb{E}_{s' \sim p} \left[\ \vs{r}(s, \vs{a}) + \gamma \vs{v}^{\nash}(s')\  \right]  \label{nashop}
\end{align}
forms a contraction mapping, where $\vs{Q} \triangleq [Q^1, \dots, Q^N]$, and $\vs{r}(s, \vs{a}) \triangleq [r^1(s, \vs{a}), \dots, r^N(s, \vs{a})]$.
The $Q$-function will eventually converge to the value received in a Nash equilibrium of the game, referred to as the \emph{Nash $Q$-value}.

\section{Mean Field MARL} \label{sec:MFQL}

The dimension of joint action $\vs{a}$ grows proportionally \emph{w.r.t.} the number of agents $N$. As all agents act strategically and evaluate simultaneously their value functions based on the joint actions, it becomes infeasible to learn the standard $Q$-function $Q^j(s,\vs{a})$. To address this issue, we factorize the $Q$-function using only the pairwise local interactions:
\begin{align}
Q^j(s, \vs{a}) &= \frac{1}{N^j} \sum_{k\in \mathcal{N}(j)} Q^j(s, a^j, a^k)~, \label{pairwise-Q}
\end{align}
\vskip -0.15in
where $\mathcal{N}(j)$ is the index set of the neighboring agents of agent $j$ with the size $N^j = |\mathcal{N}(j)|$ determined by the settings of different applications. It is worth noting that the pairwise approximation of the agent and its neighbors, while significantly reducing the complexity of the interactions among agents, still preserves global interactions between any pair of agents implicitly \cite{blume1993statistical}. Similar approaches can be found in factorization machine \cite{rendle2012factorization} and learning to rank \cite{cao2007learning}. 


\subsection{Mean Field Approximation}
The pairwise interaction $Q^j(s, a^j, a^k)$ as in Eq.~\eqref{pairwise-Q} can be approximated using the mean field theory \cite{stanley1971phase}. 
Here we consider discrete action spaces, where the action $a^j$ of agent $j$ is a discrete categorical variable represented as the one-hot encoding with each component indicating one of the $D$ possible actions: $a^j \triangleq [a^j_1,\dots,a^j_D]$. 
We calculate the \emph{mean} action $\bar{a}^j$ based on the neighborhood $\mathcal{N}(j)$ of agent $j$, and express the one-hot action $a^k$ of each neighbor $k$ in terms of the sum of $\bar{a}^j$ and a small fluctuation $\delta{a^{j,k}}$ as
\vskip -0.30in
\begin{align}
\label{eq:abar}
a^k = \bar{a}^j + \delta{a^{j,k}}, \quad\mbox{~~where~~} \bar{a}^j = \frac{1}{N^j} \sum_{k} a^k,
\end{align}
\vskip -0.225in
where $\bar{a}^j \triangleq [\bar{a}^j_1, \dots, \bar{a}^j_D]$ can be interpreted as the empirical distribution of the actions taken by agent $j$'s neighbors. By Taylor's theorem, the pairwise $Q$-function $Q^j(s, a^j, a^k)$, if twice-differentiable \emph{w.r.t.} the action $a^k$ taken by neighbor $k$, can be expended and expressed as
\vskip -0.25in
{\small
\begin{align}
&\!\!Q^j(s, \vs{a}) = \frac{1}{N^j} \sum_{k} Q^j(s, a^j, a^k) \notag\\
&= \frac{1}{N^j} \sum_{k} \bigg[ Q^j(s, a^j, \bar{a}^j) + \nabla_{\bar{a}^j} Q^j(s, a^j, \bar{a}^j) \cdot \delta{a^{j,k}} \notag\\
&\qquad\quad + \frac[1.5pt]{1}{2}\,\delta{a^{j,k}} \cdot \nabla^2_{\tilde{a}^{j,k}} Q^j(s, a^j, \tilde{a}^{j,k}) \cdot \delta{a^{j,k}} \bigg] \notag\\
&= Q^j(s, a^j, \bar{a}^j) + \nabla_{\bar{a}^j} Q^j(s, a^j, \bar{a}^j) \cdot \bigg[ \frac{1}{N^j} \sum_{k} \delta{a^{j,k}} \bigg] \notag\\
&\qquad\quad + \frac{1}{2N^j} \sum_{k} \bigg[ \delta{a^{j,k}} \cdot \nabla^2_{\tilde{a}^{j,k}} Q^j(s, a^j, \tilde{a}^{j,k}) \cdot \delta{a^{j,k}} \bigg] \label{mean-field-first} \\
&= Q^j(s, a^j, \bar{a}^j) + \frac{1}{2N^j} \sum_{k} R^j_{s, a^j}(a^k) \ \approx \ Q^j(s, a^j, \bar{a}^j)~, \label{mean-field-final}
\end{align}
}%
\vskip -0.2in
%
where $R^j_{s, a^j}(a^k) \triangleq \delta{a^{j,k}} \cdot \nabla^2_{\tilde{a}^{j,k}} Q^j(s, a^j, \tilde{a}^{j,k}) \cdot \delta{a^{j,k}}$ denotes the Taylor polynomial's remainder with $\tilde{a}^{j,k} = \bar{a}^j + \epsilon^{j,k} \delta{a^{j,k}}$ and $\epsilon^{j,k} \in [0,1]$. In Eq.~\eqref{mean-field-first}, $\sum_{k} \delta{a^k}=0$ by Eq.~\eqref{eq:abar} such that the first-order term is dropped. From the perspective of agent $j$, the action $a^k$ in the second-order remainders $R^j_{s, a^j}(a^k)$ is chosen based on the external action distribution of agent $k$, $R^j_{s, a^j}(a^k)$ is thus essentially a random variable. 
In fact, one can further prove
that the remainder $R^j_{s, a^j}(a^k)$ is bounded within a symmetric interval $[-2M,2M]$ under the mild condition of the $Q$-function $Q^j(s, a^j, a^k)$ being $M$-smooth (\emph{e.g.} the linear function); as a result, $R^j_{s, a^j}(a^k)$ acts as a small fluctuation near zero. To stay self-contained, the derivation of the bound is put in the Appendix B. 
With the assumptions of homogeneity and locality on all agents within the neighborhood, the remainders tend to cancel each other, leading to the  left term of $Q^j(s, a^j, \bar{a}^j)$ in Eq.~\eqref{mean-field-final}.

\begin{SCfigure}[][t]
\includegraphics[width=.47\columnwidth]{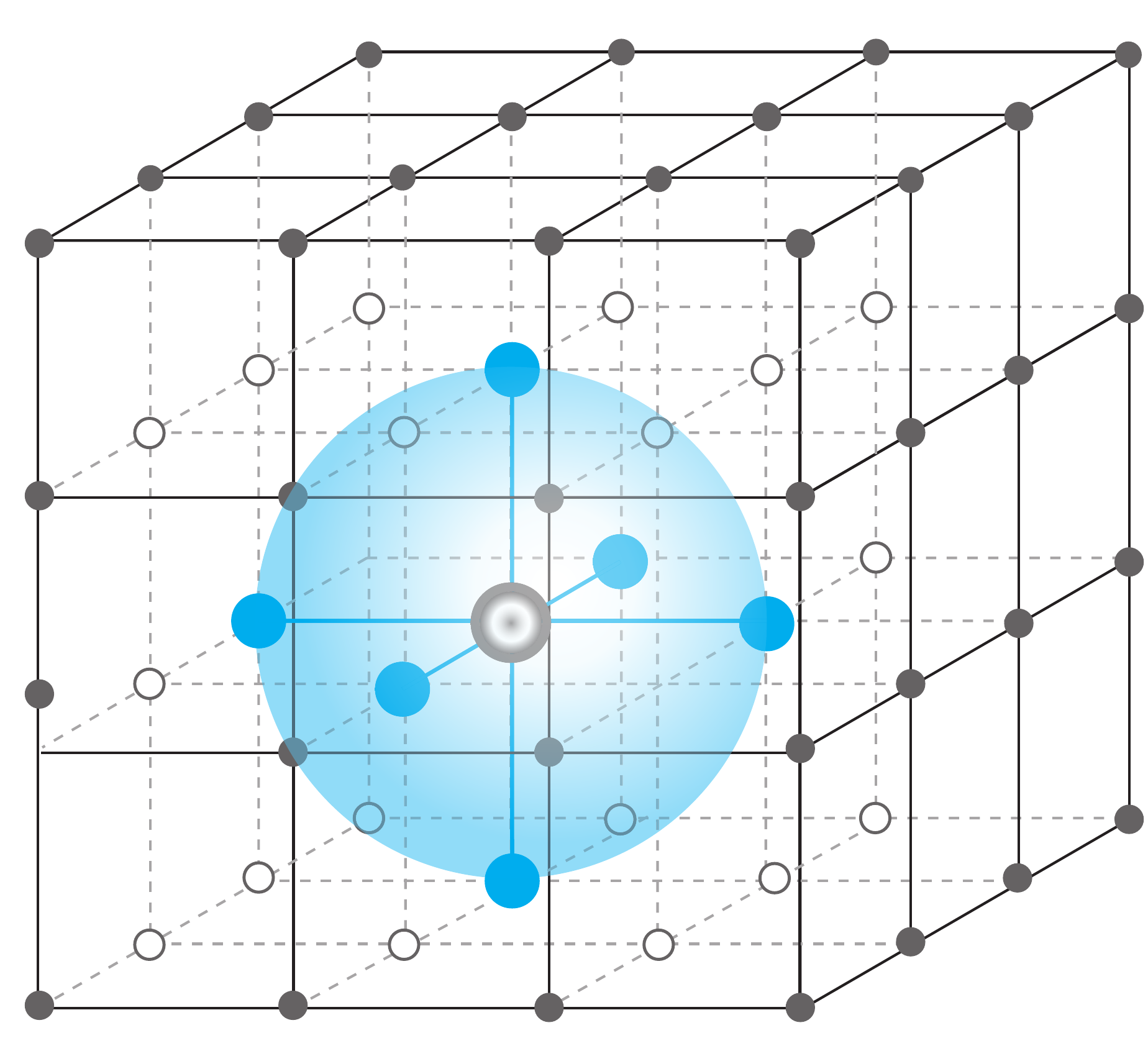}
		\caption{Mean field approximation. Each agent is represented as a node in the grid, which is only affected by the \textbf{mean} effect from its \textbf{neighbors} (the blue area). Many-agent interactions are effectively converted into two-agent interactions.}
		\label{mf}
\end{SCfigure}

As illustrated in Fig. \ref{mf}, with the mean field approximation, the pairwise interactions $Q^j(s, a^j, a^k)$ between agent $j$ and each neighboring agent $k$ are simplified as that between $j$, the \emph{central} agent, 
 and the virtual \emph{mean} agent, that is abstracted by the mean effect of all neighbors within $j$'s neighborhood. 
The interaction is thus simplified and expressed by the mean field $Q$-function $Q^j(s, a^j, \bar{a}^j)$ in Eq.~\eqref{mean-field-final}. During the learning phase, given an experience $e = \big(s, \{a^k\}, \{r^j\}, s'\big)$, the mean field $Q$-function is updated in a recurrent manner as
\vskip -0.2in
{\small
\begin{align}
Q^j_{t+1}(s, a^j, \bar{a}^j) &= (1-\alpha) Q^j_t(s, a^j, \bar{a}^j) + \alpha [ r^j + \gamma v^j_t(s') ]~,
\label{main_mfq} 
\end{align}
}%
\vskip -0.1in
where $\alpha_t$ denotes the learning rate, and $\bar{a}^j$ is the mean action of all neighbors of agent $j$ as defined in Eq.~\eqref{eq:abar}. The mean field value function $v^j_t(s')$ for agent $j$ at time $t$ in Eq.~\eqref{main_mfq} is
\vskip -0.15in 
{\small
\begin{align}
v^j_t(s') = \sum_{a^j} \pi^j_t\big(a^j | s', \bar{a}^j\big) \mathbb{E}_{\bar{a}^j(\vs{a}^{-j})\sim\vs{\pi}^{-j}_t} \Big[ Q^j_t\big( s', a^j, \bar{a}^j \big) \Big],
\label{mfv}
\end{align}
}%
\vskip -0.15in
As shown in Eqs.~\eqref{main_mfq} and \eqref{mfv}, with the mean field approximation, the MARL problem is converted into that of solving for the central agent $j$'s best response $\pi^j_t$ \emph{w.r.t.} the mean action $\bar{a}^j$ of all $j$'s neighbors, which represents the action distribution of all neighboring agents of the central agent $j$.

We introduce an iterative procedure in computing the best response $\pi^j_t$ of each agent $j$. In the stage game $\{\vs{Q}_t\}$, the mean action $\bar{a}^j$ of all $j$'s neighbors is first calculated by averaging the actions $a^k$ taken by $j$'s $N^j$ neighbors from the policies $\pi^k_t$ parametrized by their previous mean actions $\bar{a}^k_{-}$
\vskip -0.25in 
\begin{align}
\bar{a}^j = \frac{1}{N^j} \sum_{k} a^k, \  a^k\sim \pi^k_{t}(\cdot | s, \bar{a}^k_{-})~,
\label{forward_eq}
\end{align}
\vskip -0.2in
With each $\bar{a}^j$ calculated as in Eq.~\eqref{forward_eq}, the policy $\pi^j_{t}$ changes consequently due to the dependence on the current $\bar{a}^j$. The new Boltzmann policy is then determined for each $j$ that
\vskip -0.25in 
\begin{align}
\pi^j_{t}(a^j | s, \bar{a}^j) = \frac{\exp{\big( \beta Q^j_{t}(s, a^j, \bar{a}^j)\big)}}{\sum_{\substack{a^{j'}\in\mathcal{A}^j}}{\exp{\big( \beta Q^j_{t}(s, a^{j'}, \bar{a}^j)\big)}}}~.
\label{boltzman_q}
\end{align}
\vskip -0.15in

By iterating Eqs.~\eqref{forward_eq} and \eqref{boltzman_q}, the mean actions $\bar{a}^j$ and the corresponding policies $\pi^j_t$ for all agents improves alternatively. In spite of lacking an intuitive impression of being stationary, in the following subsections, we will show that the mean action $\bar{a}^j$ will be equilibrated at an unique point after several iterations, and hence the policy $\pi^j_t$ converges.

To distinguish from the Nash value function $\vs{v}^{\nash}(s)$ in Eq.~\eqref{nashop}, we denote the mean field value function in Eq.~\eqref{mfv} as $\vs{v}^{\mf}(s) \triangleq [v^{1}(s), \dots, v^{N}(s)]$.
With $\vs{v}^{\mf}$ assembled, we now define the mean field operator $\mathcal{H}^{\mf}$ in the form of
\vskip -0.25in 
\begin{align}
\mathcal{H}^{\mf} \vs{Q}(s, \vs{a}) = \mathbb{E}_{s' \sim p} \left[\ \vs{r}(s, \vs{a}) + \gamma \vs{v}^{\mf}(s')\ \right]. \label{mfop}
\end{align}
\vskip -0.10in
In fact, we can prove that $\mathcal{H}^{\mf}$ forms a contraction mapping; that is, one updates $\vs{Q}$ by iteratively applying the mean field operator $\mathcal{H}^{\mf}$, the mean field $Q$-function will eventually converge to the Nash $Q$-value under certain assumptions.

\subsection{Implementation}
We can implement the mean field $Q$-function in Eq.~\eqref{mean-field-final} by universal function approximators such as neural networks, where the $Q$-function is parameterized with the weights $\phi$. The update rule in Eq.~\eqref{main_mfq} can be reformulated as weights adjustment. 
For off-policy learning, we exploit either standard $Q$-learning \cite{watkins1992q} for discrete action spaces or DPG \cite{silver2014deterministic} for continuous action spaces.
Here we focus on the former, which we call MF-$Q$.

In MF-$Q$, agent $j$ is trained by minimizing the loss function
\vskip -0.25in
\begin{align*}
\mathcal{L}(\phi^j) = \big(y^j - Q_{\phi^j}(s, a^j, \bar{a}^j)\big)^2,
\end{align*}
\vskip -0.15in
where $y^j = r^j+\gamma\, v^{\mf}_{\phi^j_{-}}(s')$ is the target mean field value calculated with the weights $\phi^j_{-}$. Differentiating $\mathcal{L}(\phi^j)$  gives
\vskip -0.25in
{\small
\begin{align}
\label{mfq-differentiate}
\nabla_{\phi^j} \mathcal{L}(\phi^j)=\Big( y^j - {Q_{\phi^j}(s, a^j, \bar{a}^j)}\Big) \nabla_{\phi^j}{Q_{\phi^j}(s, a^j, \bar{a}^j)}~,
\end{align}
}%
\vskip -0.15in
which enables the gradient-based optimizers for training.

Instead of setting up Boltzmann policy using the $Q$-function as in MF-$Q$, we can explicitly model the policy by neural networks with the weights $\theta$, which leads to the on-policy actor-critic method \cite{NIPS1999_1786} that we call MF-AC. The policy network $\pi_{\theta^j}$, \emph{i.e.} the actor, of MF-AC is trained by the sampled policy gradient:
\vskip -0.275in
\begin{align*}
\nabla_{\theta^j}\mathcal{J}(\theta^j) \approx \nabla_{\theta^j} \log\pi_{\theta^j}(s)Q_{\phi^j}(s,a^j, \bar{a}^j)\Big|_{a=\pi_{\theta^j}(s)}~.
\end{align*}
\vskip -0.175in
The critic of MF-AC follows the same setting for MF-$Q$ with Eq.~\eqref{mfq-differentiate}.
During the training of MF-AC, one needs to alternatively update $\phi$ and $\theta$ until convergence. We illustrate the MF-$Q$ iterations in Fig.~\ref{ising_mfq}, and present the pesudocode for both MF-$Q$ and MF-AC in Appendix A. 

\subsection{Proof of Convergence}\label{sec:proof-conv}

We now prove the convergence of $\vs{Q}_t \triangleq [Q^1_t, \dots, Q^N_t]$ to the Nash $Q$-value $\vs{Q}_* = [Q^1_*, \dots, Q^N_*]$ as the iterations of MF-$Q$ is applied. The proof is presented by showing that the mean field operator $\mathcal{H}^{\mf}$ in Eq.~\eqref{mfop} forms a contraction mapping with the fixed point at $\vs{Q}_*$ under the main assumptions. We start from introducing the assumptions:

\begin{assumption}\label{tableassum}
Each action-value pair is visited infinitely often, and the reward is bounded by some constant $K$.
\end{assumption}

\begin{assumption}\label{glieassum}
Agent's policy is Greedy in the Limit with Infinite Exploration (GLIE). In the case with the Boltzmann policy, the policy becomes greedy \emph{w.r.t.} the $Q$-function in the limit as the temperature decays asymptotically to zero. 
\end{assumption}

\begin{assumption}\label{2nasheq}
For each stage game $[Q_t^1(s), ..., Q_t^N(s)]$ at time $t$ and in state $s$ in training, for all $t$, $s$, $j\in \{1,\dots,N\}$, 
the Nash equilibrium $\vs{\pi}_* = [\pi^1_*, \dots, \pi^N_*]$ is recognized either as 1) the \emph{global optimum} or 2) a \emph{saddle point} expressed as: 
\vspace{-3mm}
\begin{itemize}[leftmargin=10pt]\justifying \setlength\itemsep{.01em}
\item[1.] $\mathbb{E}_{\vs{\pi}_*} [Q_t^j(s)] \ge \mathbb{E}_{\vs{\pi}} [Q_t^j(s)],\ \forall\vs{\pi} \in \Omega\big(\prod_k \mathcal{A}^k\big)$;
\item[2.] $\mathbb{E}_{\vs{\pi}_*} [Q_t^j(s)] \ge \mathbb{E}_{\pi^j}\mathbb{E}_{\vs{\pi}^{-j}_*} [Q_t^j(s)],\ \forall\pi^j \in \Omega\big(\mathcal{A}^j\big)$ and\\$\mathbb{E}_{\vs{\pi}_*} [Q_t^j(s)] \le \mathbb{E}_{\pi^j_*}\mathbb{E}_{\vs{\pi}^{-j}} [Q_t^j(s)],\ \forall\vs{\pi}^{-j} \in \Omega\big(\prod_{k\ne j}\mathcal{A}^k\big)$.
\end{itemize}	
\vspace{-3mm}
\end{assumption}

	
\begin{figure}[t]
\vskip -0.1in
\centering
\includegraphics[width=\columnwidth]{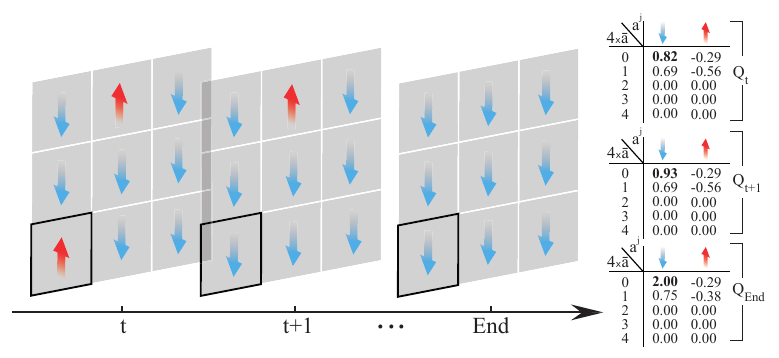}
\vskip -0.15in
\caption{MF-$Q$ iterations on a $3\times 3$ stateless toy example. The goal is to coordinate the agents to an agreed direction.
Each agent has two choices of actions: \emph{up} $\uparrow$ or \emph{down} $\downarrow$. 
The reward of each agent's staying in the same direction as its [$0,1,2,3,4$] neighbors are [$-2.0, -1.0, 0.0, 1.0, 2.0$], respectively. The neighbors are specified by the four directions on the grid with cyclic structure on all directions, \emph{e.g.} the first row and the third row are adjacent. 
The reward for the highlighted agent $j$ on the bottom left at time $t+1$ is $2.0$, as all neighboring agents stay down in the same time. 
We listed the Q-tables for agent $j$ at three time steps where $\bar{a}^j$ is the percentage of neighboring ups. Following Eq.~\ref{main_mfq}, we have $Q^j_{t+1}(\uparrow, \bar{a}^j=0) = Q^j_t(\uparrow, \bar{a}^j=0) + \alpha [ r^j -  Q^j_t(\uparrow, \bar{a}^j=0)] = 0.82 + 0.1\times (2.0-0.82) = 0.93$.
The rightmost plot shows the convergent scenario where the $Q$-value of staying down is $2.0$, which is the largest reward in the environment. 
}
\label{ising_mfq}
\vskip -0.in
\end{figure}

Note that Assumption \ref{2nasheq} imposes a strong constraint on every single stage game encountered in training. In practice, however, we find this constraint appears not to be a necessary condition for the learning algorithm to converge. This is in line with the empirical findings in \citet{hu2003nash}.

Our proof is also built upon the two lemmas as follows:
\begin{lemma}\label{nashq}
Under Assumption \ref{2nasheq},
the Nash operator $\mathcal{H}^{\nash}$ in Eq.~\eqref{nashop}
forms a contraction mapping on the complete metric space from $\mathcal{Q}$ to $\mathcal{Q}$ with the fixed point being the Nash $Q$-value of the entire game, \emph{i.e.}  $\mathcal{H}^{\nash}_t \vs{Q}_* = \vs{Q}_*$.
\end{lemma}
\vspace{-3mm}
\begin{proof}
See Theorem 17 in \citet{hu2003nash}.	
\end{proof}

%
%

\begin{lemma}\label{fundamental}
The random process $\{\Delta_t\}$ defined in $\mathbb{R}$ as
\begin{equation}
\Delta_{t+1}(x) = (1 - \alpha_t(x))\Delta_{t}(x) + \alpha_t(x)F_t(x)
\label{lemma1}
\end{equation}
converges to zero with probability $1$ (\emph{w.p.$1$}) when
\begin{itemize}[leftmargin=10pt]\justifying \setlength\itemsep{.01em}
\vspace{-3mm}
\item[1.] $0\leq \alpha_t(x) \leq 1$, $\sum_t \alpha_t(x) = \infty$, $\sum_t \alpha_t^2(x) < \infty$;
\item[2.] $x \in \mathcal{X}$, the set of possible states, and $|\mathcal{X}| < \infty$;
\item[3.] $\| \mathbb{E}[F_t(x) | \mathcal{F}_t ] \|_{W} \leq \gamma \|\Delta_t\|_W + c_t$, where $\gamma \in [0,1)$ and $c_t$ converges to zero \emph{w.p.$1$};
\item[4.] $\mathbf{var}[F_t(x) | \mathcal{F}_t ] \leq K(1+\|\Delta_t\|_{W}^2)$ with constant $K>0$.
\end{itemize}	
\vspace{-3mm}
Here $\mathcal{F}_t$ denotes the filtration of an increasing sequence of $\sigma$-fields including the history of processes; $\alpha_t, \Delta_t, F_t \in \mathcal{F}_t$ and $\|\cdot\|_W$ is a weighted maximum norm \cite{bertsekas2012weighted}.
\end{lemma}
\vspace{-3mm}
\begin{proof}
See Theorem 1 in \citet{jaakkola1994convergence} and Corollary 5 \citet{szepesvari1999unified} for detailed derivation.
We include it here to stay self-contained. 
\end{proof}
\vspace{-2mm}

By subtracting $\vs{Q}_*(s, \vs{a})$ on both sides of Eq.~\eqref{main_mfq}, we present the relation from the comparison with Eq.~\eqref{lemma1} such that
\vskip -0.30in
\begin{align}
\vs{\Delta}_t(x) &= \vs{Q}_t(s, \vs{a}) - \vs{Q}_*(s, \vs{a}), \notag\\
\vs{F}_t(x) &= \vs{r}_t + \gamma \vs{v}^{\mf}_t(s_{t+1}) - \vs{Q}_*(s_t, \vs{a}_t), \label{eq:ft}
\end{align}
\vskip -0.15in
where $x \triangleq (s_t, \vs{a}_t)$ denotes the visited state-action pair at time $t$. In Eq.~\eqref{lemma1}, $\alpha(t)$ is interpreted as the learning rate with $\alpha_t(s', \vs{a}') = 0$ for any $(s', \vs{a}') \neq (s_t, \vs{a}_t)$; this is because that each agent only updates the $Q$-function with the state $s_t$ and actions $\vs{a}_t$ visited at time $t$. Lemma \ref{fundamental} suggests $\Delta_t(x)$'s convergence to zero, which means, if it holds, the sequence of $Q$'s will asymptotically tend to the Nash $Q$-value $\vs{Q}_*$. 
One last piece to establish the main theorem is the below:

\begin{restatable}{proposition}{bolzproposition}
\label{bolz}
Let the metric space be $\mathbb{R}^N$ and the metric be $d(\vs{a}, \vs{b}) = \sum_j{| a^j - b^j |}$, for $\vs{a}=[a^j]_1^N, \vs{b}=[b^j]_1^N$. If the $Q$-function is $K$-Lipschitz continuous \emph{w.r.t.}  $a^j$, then
the operator  $\mathcal{B}(a^j) \triangleq \pi^j(a^j | s, \bar{a}^j) $ in Eq.~\eqref{boltzman_q}
forms a contraction mapping under sufficiently low temperature $\beta$.
\end{restatable}

\begin{proof}
See details in Appendix D due to the space limit.
\end{proof}
\begin{theorem}


In a finite-state stochastic game, the $\vs{Q}_t$ values computed by the update rule of MF-$Q$ in Eq.~\eqref{main_mfq} converges to the Nash $Q$-value $\vs{Q}_* = [Q_*^{1}, \dots , Q_*^{N}]$,  if Assumptions \ref{tableassum}, \ref{glieassum} \& \ref{2nasheq}, and Lemma \ref{fundamental}'s first and second conditions are met.
\end{theorem}
\vspace{-3mm}
\begin{proof}
Let $\mathcal{F}_t$ denote the $\sigma$-field generated by all random variables in the history of the stochastic game up to time $t$: $(s_t, \alpha_t, \vs{a}_t, r_{t-1}, ..., s_1, \alpha_1, \vs{a}_1, \vs{Q}_0)$. Note that $\vs{Q}_t$ is a random variable derived from the historical trajectory up to time $t$. Given the fact that all $\vs{Q}_{\tau}$ with $\tau < t$ are $\mathcal{F}_t$-measurable, both $\vs{\Delta}_t$ and $\vs{F}_{t-1}$ are therefore also $\mathcal{F}_t$-measurable, which satisfies the measurability condition of Lemma \ref{fundamental}.

To apply Lemma \ref{fundamental}, we need to show that the mean field operator $\mathcal{H}^{\mf}$ meets Lemma \ref{fundamental}'s third and fourth conditions. For Lemma \ref{fundamental}'s third condition, we begin with Eq. \eqref{eq:ft} that
\begin{align}
\vs{F}_t(s_t, \vs{a}_t) &= \vs{r}_t + \gamma \vs{v}^{\mf}_t(s_{t+1}) - \vs{Q}_*(s_t, \vs{a}_t) \nonumber \\
&= \vs{r}_t + \gamma \vs{v}^{\nash}_t(s_{t+1}) - \vs{Q}_*(s_t, \vs{a}_t) \nonumber \\
&\quad + \gamma[\vs{v}^{\mf}_t(s_{t+1}) - \vs{v}^{\nash}_t(s_{t+1}) ] \nonumber \\
&= \left[\vs{r}_t + \gamma \vs{v}^{\nash}_t(s_{t+1}) - \vs{Q}_*(s_t, \vs{a}_t) \right] + C_t(s_t, \vs{a}_t) \nonumber \\
&= \vs{F}_t^{\nash}(s_t, \vs{a}_t) + \vs{C}_t(s_t, \vs{a}_t).
\label{eq:fnash}
\end{align}
Note the fact that $\vs{F}_t^{\nash}$ in Eq. \eqref{eq:fnash} is essentially the $\vs{F}_t$ in Lemma \ref{fundamental} in proving the convergence of the Nash $Q$-learning algorithm.
From Lemma \ref{nashq}, it is straightforward to show that $\vs{F}_t^{\nash}$ forms a contraction mapping with the norm $\|\cdot\|_{\infty}$ being the maximum norm on $\vs{a}$. We thus have for all $t$ that 
\begin{align*}
\| \mathbb{E}[\vs{F}_t^{\nash}(s_t, \vs{a}_t) | \mathcal{F}_t]\|_{\infty} \leq  \gamma \|\vs{Q}_t- \vs{Q}_*\|_{\infty} = \gamma\|\vs{\Delta}_t\|_{\infty}.
\end{align*}
\vspace{-7mm}


In meeting the third condition, we obtain from Eq.~\eqref{eq:fnash} that
\vspace{-1mm}
{\small
\begin{align}
\| \mathbb{E}[\vs{F}_t(s_t, \vs{a}_t) | \mathcal{F}_t ] \|_{\infty} &\leq \|\vs{F}_t^{\nash}(s_t, \vs{a}_t) | \mathcal{F}_t\|_{\infty}  + \|\vs{C}_t(s_t, \vs{a}_t) | \mathcal{F}_t\|_{\infty}  \nonumber \\
 &\leq \gamma\|\vs{\Delta}_t\|_{\infty} + \|\vs{C}_t(s_t, \vs{a}_t) | \mathcal{F}_t\|_{\infty} .
\end{align}
}%
We are left to prove that $c_t =  \|\vs{C}_t(s_t, \vs{a}_t) | \mathcal{F}_t\|$ converges to zero \emph{w.p.$1$}. With Assumption \ref{2nasheq}, for each stage game, all the globally optimal equilibrium(s) share the same Nash value, so does the saddle-point equilibrium(s). Each of the two following results is essentially associated with one of the two mutually exclusive scenarios in Assumption \ref{2nasheq}:
\vspace{-3mm}
\begin{itemize}[leftmargin=10pt]\justifying \setlength\itemsep{.01em}
\item[1.] For globally optimal equilibriums, all players obtain the joint maximum values that are unique and identical for all equilibriums according to the definition;
\item[2.] Suppose that the stage game $\{\vs{Q}_t\}$ has two saddle-point equilibriums, $\vs{\pi}$ and $\vs{\rho}$. It holds for agent $j$ that 
\vspace{-2mm}
\begin{align}
\mathbb{E}_{\pi^j}\mathbb{E}_{\vs{\pi}^{-j}} [Q_t^j(s)] &\ge \mathbb{E}_{\rho^j}\mathbb{E}_{\vs{\pi}^{-j}} [Q_t^j(s)], \nonumber \\
\mathbb{E}_{\rho^j}\mathbb{E}_{\vs{\rho}^{-j}} [Q_t^j(s)] &\le \mathbb{E}_{\rho^j}\mathbb{E}_{\vs{\pi}^{-j}} [Q_t^j(s)] \nonumber.
\end{align}
\vskip -0.15in
By combing the above inequalities, we obtain 
\vskip -0.3in
\begin{align*}
\mathbb{E}_{\pi^j}\mathbb{E}_{\vs{\pi}^{-j}} [Q_t^j(s)] \ge \mathbb{E}_{\rho^j}\mathbb{E}_{\vs{\rho}^{-j}} [Q_t^j(s)].
\end{align*}
\vskip -0.15in
By the definition of saddle points, the above inequality still holds by reversing the order of $\vs{\pi}$ and $\vs{\rho}$; hence, the equilibriums for agent $i$ at both saddle points are the same such that $\mathbb{E}_{\pi^j}\mathbb{E}_{\vs{\pi}^{-j}} [Q_t^j(s)] = \mathbb{E}_{\rho^j}\mathbb{E}_{\vs{\rho}^{-j}} [Q_t^j(s)]$.
\end{itemize}	
\vspace{-2mm}

%

Given Proposition \ref{bolz} that the policy based on the mean field $Q$-function forms a contraction mapping, and that all optimal/saddle points share the same Nash value in each stage game, with the homogeneity of agents, $\vs{v}^{\mf}$ will asymptotically converges to $\vs{v}^{\nash}$, the third condition is thus satisfied.

For the fourth condition, we exploit the conclusion that is proved above that $\mathcal{H}^{\mf}$ forms a contraction mapping, \emph{i.e.} $\mathcal{H}^{\mf}\vs{Q}_* = \vs{Q}_*$, and it follows that
{
\vskip -0.32in
\begin{align}
\mathbf{var}&[\vs{F}_t(s_t, \vs{a}_t) | \mathcal{F}_t ] =  \mathbb{E}[(\vs{r}_t + \gamma \vs{v}^{\mf}_t(s_{t+1}) - \vs{Q}_*(s_t, \vs{a}_t) )^2] \nonumber \\
&=  \mathbb{E}[(\vs{r}_t + \gamma \vs{v}^{\mf}_t(s_{t+1}) - \mathcal{H}^{\mf}(\vs{Q}_*))^2] \nonumber \\
&= \mathbf{var}[\vs{r}_t + \gamma \vs{v}^{\mf}_t(s_{t+1}) | \mathcal{F}_t] \nonumber \\
&\leq K(1+\|\vs{\Delta}_t\|_{W}^2). \label{eq:finalconverge}
\end{align}
\vskip -0.15in
}%

In the last step of Eq.~\eqref{eq:finalconverge}, we employ Assumption 1 that the reward $\vs{r}_t$ is always bounded by some constant.
Finally, with all conditions met, it follows Lemma \ref{fundamental} that $\vs{\Delta}_t$ converges to zero \emph{w.p.$1$}, \emph{i.e.} $\vs{Q}_t$ converges to $\vs{Q}_*$ \emph{w.p.$1$}.
\end{proof}
\vskip -0.05in

Apart from being convergent to the Nash $Q$-value, MF-$Q$ is also \emph{Rational} \cite{bowling2001rational, bowling2002multiagent}. We leave the corresponding discussion in Appendix D for details.
\begin{figure*}[t]
	\vskip -0.1in
	\begin{subfigure}[b]{.33\linewidth}
		\centering
		\includegraphics[width=\columnwidth]{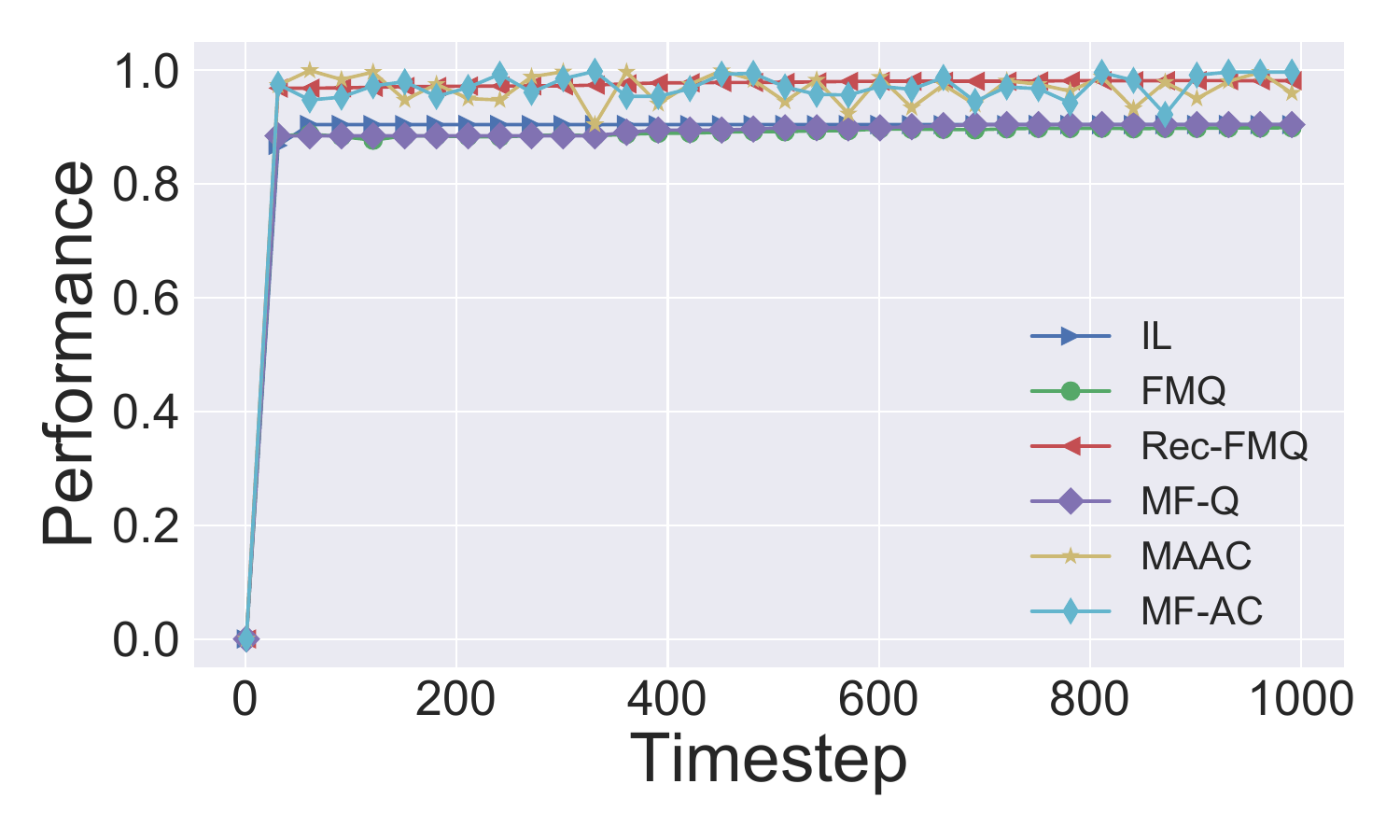}
		\vskip -0.1in
		\caption{$N = 100$}
		\label{subfig:gsd_100}
	\end{subfigure}%
	\begin{subfigure}[b]{.33\linewidth}
		\centering
		\includegraphics[width=\columnwidth]{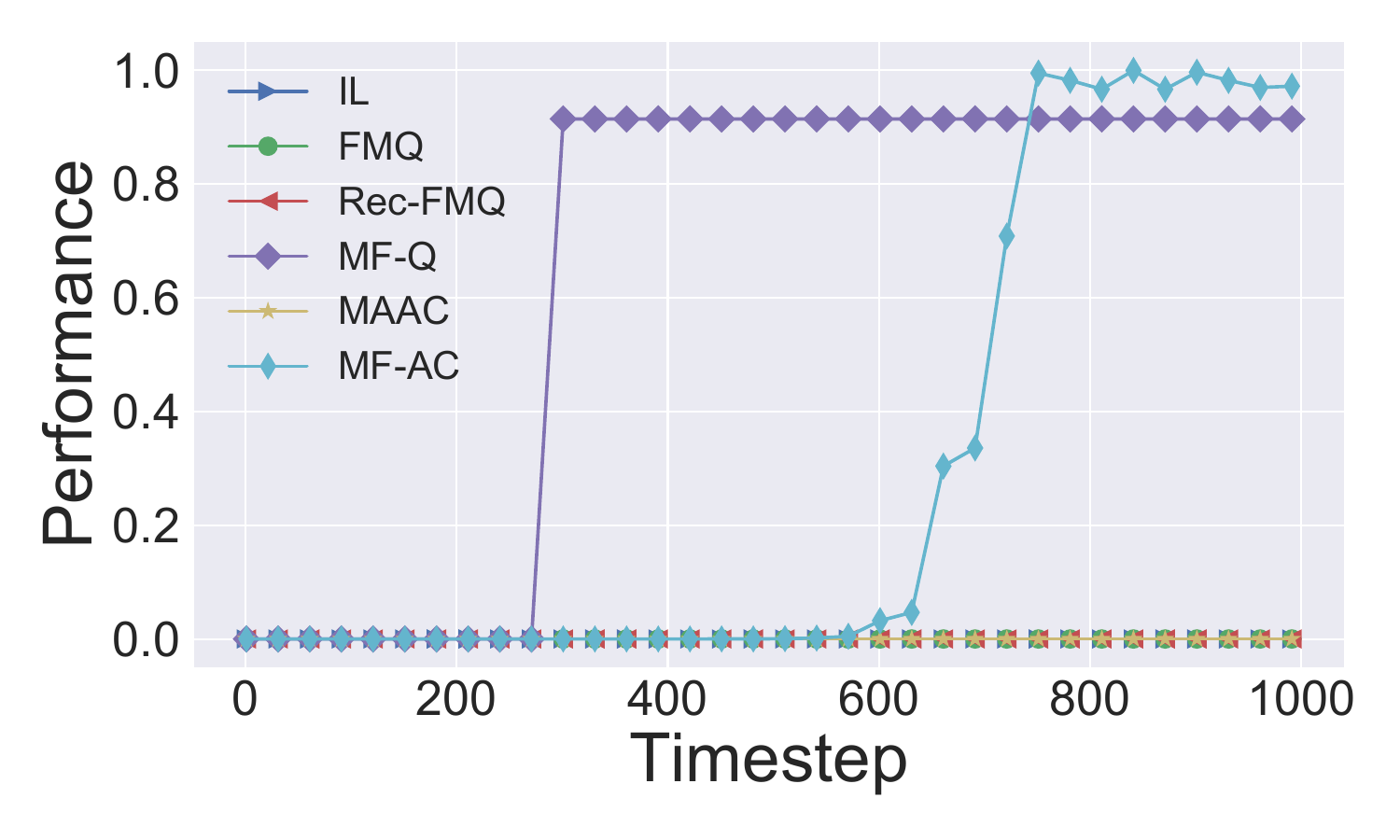}
		\vskip -0.1in
		\caption{$N = 500$}
		\label{subfig:gsd_500}
	\end{subfigure}
	\begin{subfigure}[b]{.33\linewidth}
		\centering
		\includegraphics[width=\columnwidth]{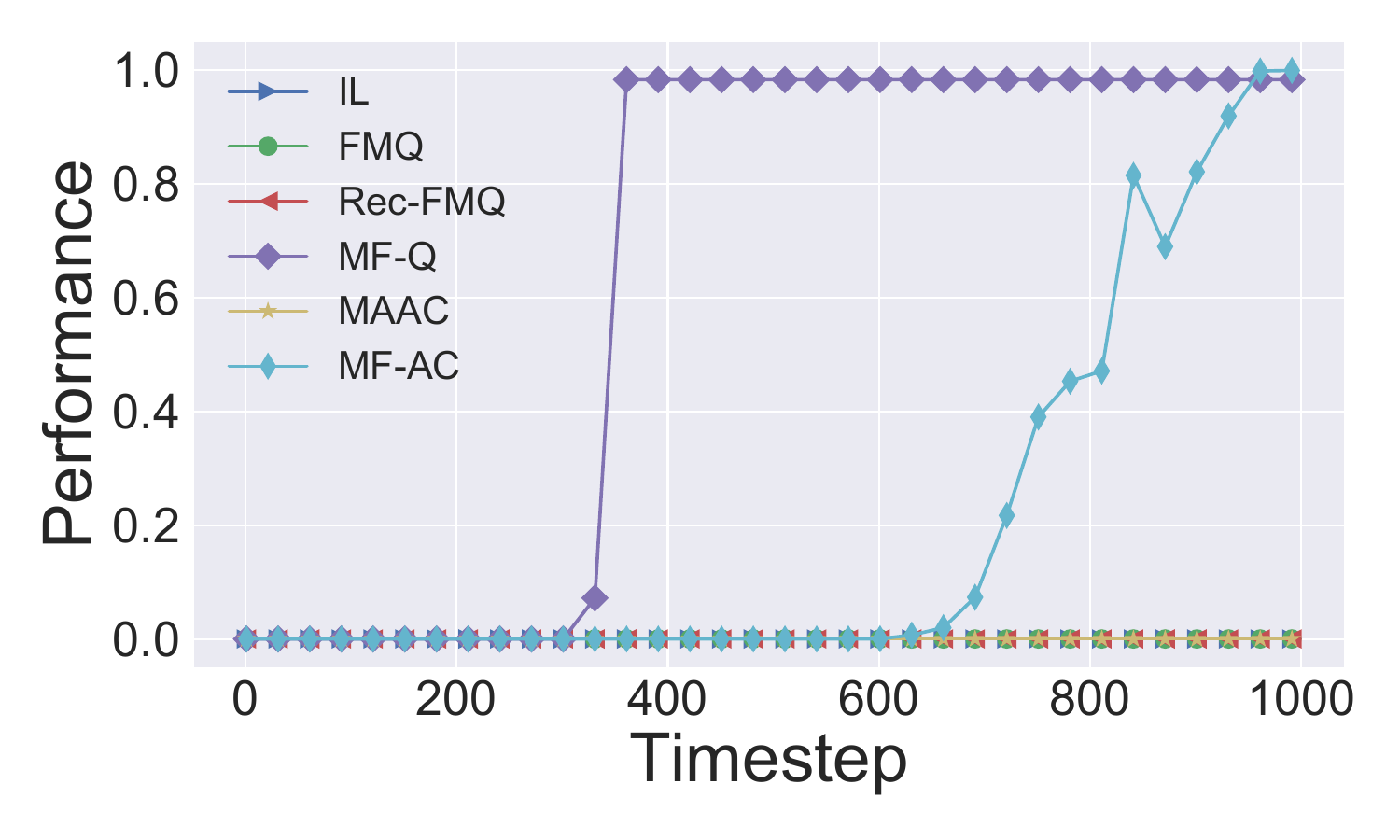}
		\vskip -0.1in
		\caption{$N = 1000$}
		\label{subfig:gsd_1000}
	\end{subfigure}
\vskip -0.1in
	\caption{Learning with $N$ agents in the GS environment with $\mu=400$ and $\sigma=200$.}
		\label{fig:gsd}
\end{figure*}

\section{Related Work}

We continue our discussion on related work from Introduction
and make comparisons with existing techniques in a greater scope. Our work follows the same direction as \citet{littman1994markov,hu2003nash,bowling2002multiagent} on adapting a Stochastic Game \cite{van1981stochastic} into the MARL formulation. Specifically, \citet{littman1994markov} addressed two-player zero-sum stochastic games by introducing a  ``minimax'' operator in $Q$-learning, whereas 
\citet{hu2003nash} extended it to the general-sum case by learning a Nash equilibrium in each stage game and considering a mixed strategy.  
Nash-Q learning is guaranteed to converge to Nash strategies under the (strong) assumption that there exists an equilibrium for every stage game. In the situation where agents can be identified as either "friends" or "foes"  \cite{littman2001friend}, one can simply solve it by alternating between fully
cooperative and zero-sum learning.
Considering the convergence speed, \citet{littman2005polynomial} and \citet{DBLP:conf/uai/CoteL08} draw on the \emph{folk theorem} and acquired a polynomial-time Nash equilibrium algorithm for repeated stochastic games, while \citet{bowling2002multiagent} tried varying the learning rate to improve the convergence.

The recent treatment of MARL was using deep neural networks as the function approximator.  
In addressing the non-stationary issue in MARL, various solutions have been proposed including 
neural-based opponent modeling \cite{DBLP:conf/icml/HeB16},  
policy parameters sharing \cite{gupta2017cooperative}, \emph{etc}.
Researchers have also adopted the paradigm of \emph{centralized training with decentralized execution} for multi-agent policy-gradient learning: BICNET \cite{DBLP:journals/corr/PengYWYTLW17}, COMA \cite{comafoerster} and MADDPG \cite{lowe2017multi}, which  allows the centralized critic $Q$-function to be trained with the actions of other agents, while the actor needs only local observation  to optimize agent's policy. 

The above MARL approaches limit their studies mostly to tens of agents. As the number of agents grows larger, not only the input space of $Q$ grows exponentially, but most critically, the accumulated noises by the exploratory
actions of other agents make the $Q$-function learning no longer feasible. Our work addresses the issue by employing the mean field approximation \cite{stanley1971phase} over the joint action space. The parameters of the $Q$-function is independent of the number of agents as it transforms multiple agents interactions into two entities interactions (single agent \emph{v.s.} the distribution of the neighboring agents). This would  effectively alleviate the problem of the  exploratory noise \cite{colby2015counterfactual} caused by many other agents,  and allow each agent to determine which actions are beneficial to itself.


Our work is also closely related to the recent development of mean field games (MFG) 
\cite{lasry2007mean,huang2006large,weintraub2006oblivious}. 
MFG studies population behaviors resulting from the aggregations of decisions taken from individuals. Mathematically, the dynamics are governed by a set of two stochastic differential equations that model the backward dynamics of individual's value function, and the forward dynamics of the aggregate distribution of agent population. Despite that the backward equation equivalently describes what Bellman equation indicates in the MDP, the primarily goal for MFG is rather for a model-based planning and to infer the movements of the individual density through time. The mean field approximation \cite{stanley1971phase} in also employed in physics, but our work is different in that we focus on a model-free solution of learning optimal actions when the dynamics of the system and the reward function are unknown. 
Very recently, \citet{DBLP:journals/corr/abs-1711-03156} built a connection between MFG and reinforcement learning. Their focus is, however, on the inverse RL in order to learn both the reward function and the forward dynamics of the MFG from the policy data, whereas our goal is to form a computable $Q$-learning algorithm under the framework of temporal difference learning.

\section{Experiments} 
We analyze and evaluate our algorithms in three different scenarios, including two stage games: the Gaussian Squeeze and the Ising Model, and the \emph{mixed cooperative-competitive} battle game. 

\subsection{Gaussian Squeeze}

\textbf{Environment.}\label{subsubsec:gaussian_environments}
In the Gaussian Squeeze  (GS) task \citep{holmesparker2014exploiting}, $N$ homogeneous agents determine their individual action $a^j$ to jointly optimize the most appropriate summation $x=\sum_{j=1}^{N}{a^j}$. Each agent has 10 action choices --  integers $0$ to $9$. The system objective is defined as $G(x)=xe^{\frac{-(x-\mu)^2}{\sigma^2}}$, where $\mu$ and $\sigma$ are the pre-defined mean and variance of the system. In the scenario of traffic congestion, each agent is one traffic controller trying to send $a^j$ vehicles into the main road. Controllers are expected to coordinate with each other to make the full use of the main route while avoiding congestions.    The goal of each agent is to learn to allocate system resources efficiently, avoiding either over-use or under-use. 
The GS problem here sits ideally as an ablation study on the impact of multi-agent exploratory noises toward the learning \cite{colby2015counterfactual}.



\textbf{Model Settings.} 
We implement MF-$Q$ and MF-AC following the framework of centralized training (shared critic) with decentralized execution (independent actor). We compare against 4 baseline models: (1) Independent Learner (IL), a traditional $Q$-Learning algorithm that does not consider the actions performed by other agents; (2) Frequency Maximum $Q$-value (FMQ) \cite{Kapetanakis:2002:RLC:777092.777145}, a modified IL which increases the $Q$-values of actions that frequently produced good rewards in the past; (3) Recursive Frequency Maximum $Q$-value (Rec-FMQ) \cite{matignon2012independent}, an improved version of FMQ that recursively computes the occurrence frequency to evaluate and then choose actions; (4) Multi-agent Actor-Critic (MAAC), a variant of MADDPG architecture  for the discrete action space (see Eq. (4) in \citet{DBLP:conf/nips/LoweWTHAM17}). All models use the multilayer perception as the function approximator. The detailed settings of the implementation are in the Appendix C.1. 

\begin{figure}[t]
\vskip -0.05in
\centering
\includegraphics[width=.75\columnwidth]{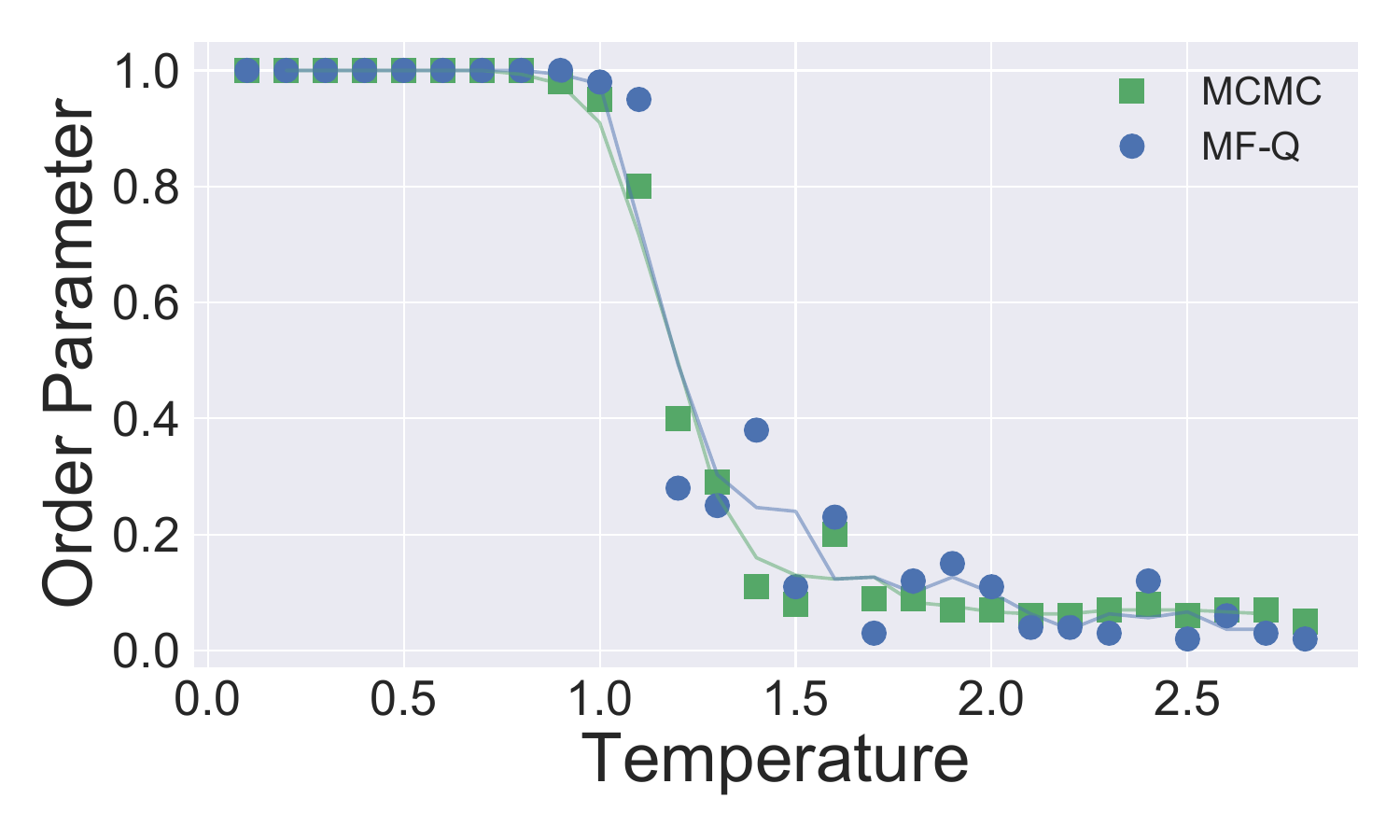}
\vskip -0.15in
\caption{The \emph{order parameter} at equilibrium \emph{v.s.} temperature in the Ising model with $20\times 20$ grid.}\label{fig:OP_T}
	\begin{subfigure}[b]{.49\linewidth}
		\centering
		\includegraphics[width=\columnwidth]{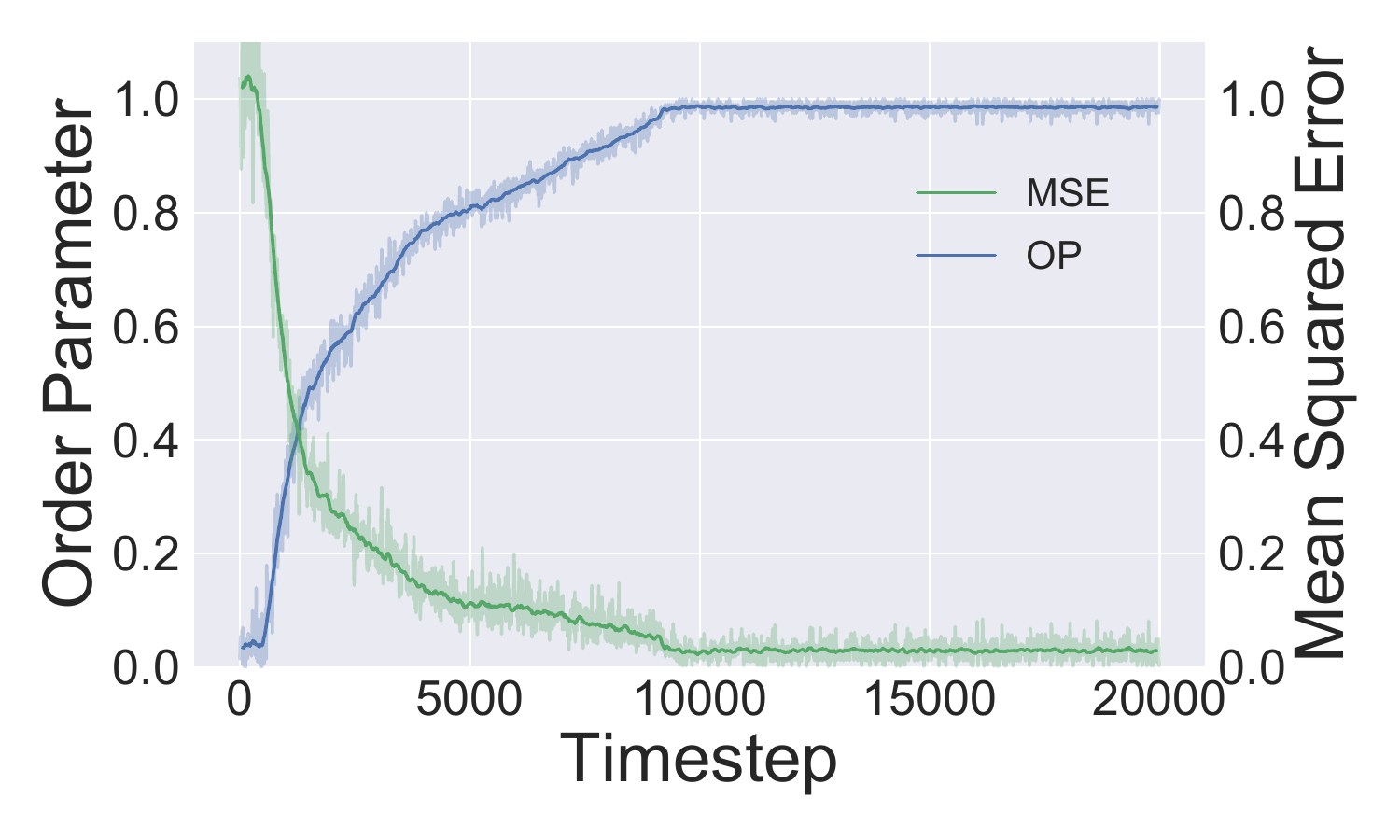}
		\vskip -0.1in
		\caption{$\tau = 0.8$}
		\label{subfig:mfq_400_0_8_mse}
	\end{subfigure}
	\begin{subfigure}[b]{.49\linewidth}
		\centering
		\includegraphics[width=\columnwidth]{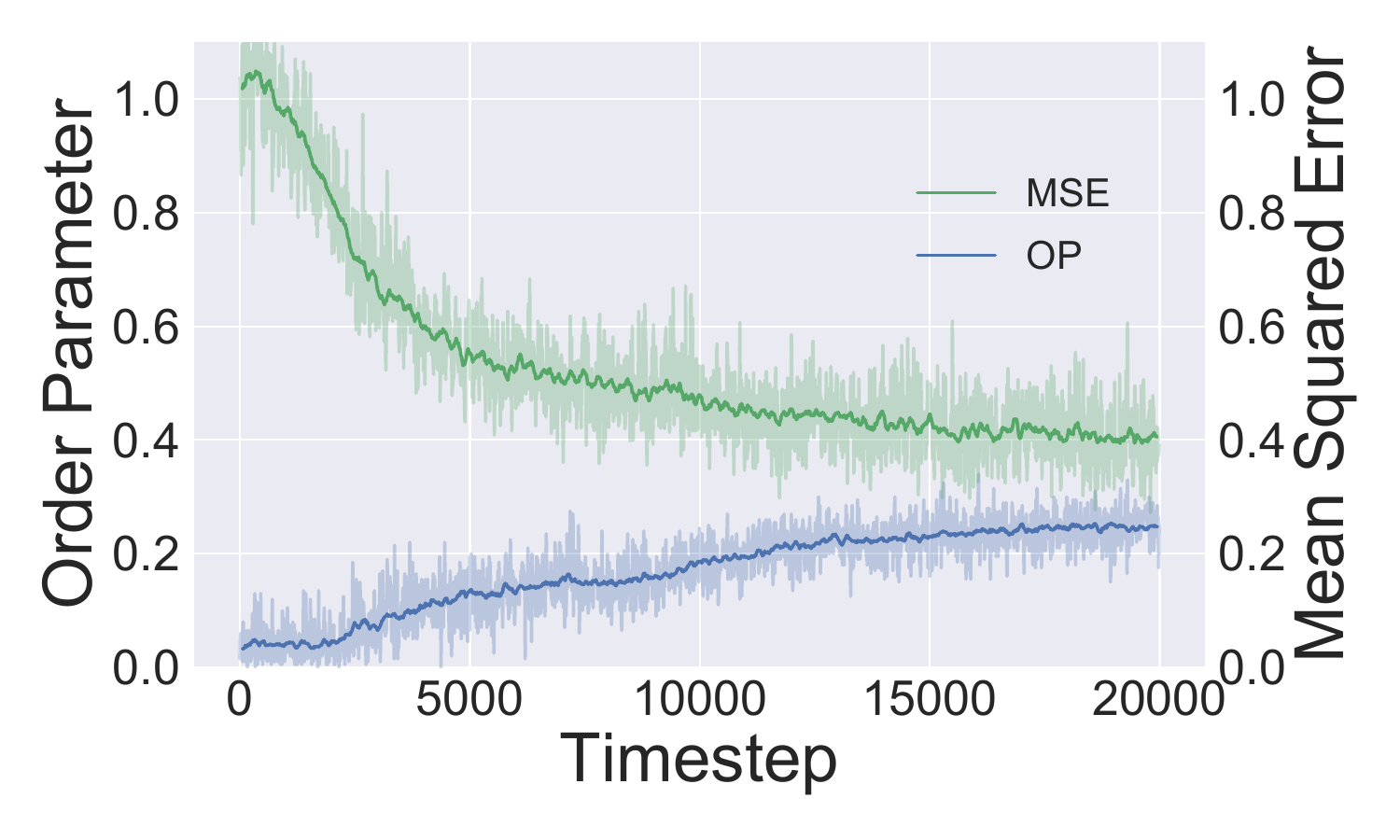}
		\vskip -0.1in
		\caption{$\tau = 1.2$}
		\label{subfig:mfq_400_1_2_mse}
	\end{subfigure}
	\vskip -0.1in
	\caption{Training performance of MF-$Q$ in the Ising model with $20\times 20$ grid.}\label{fig:mfq_mse}
\vskip -0.0in
\end{figure}
\textbf{Results.}
Figure.~\ref{fig:gsd} illustrates the results for the GS environment of $\mu=400$ and $\sigma=200$ with three different numbers of agents ($N=100, 500, 1000$) that stand for $3$ levels of congestions. 
In the smallest GS setting of Fig.~\ref{subfig:gsd_100}, all models show excellent performance. As the agent number increases, Figs.~\ref{subfig:gsd_500} and \ref{subfig:gsd_1000} show MF-$Q$ and MF-AC's  capabilities of learning the optimal allocation effectively after a few iterations, whereas all four baselines fail to learn at all. We believe this advantage is due to the awareness of other agents' actions under the mean field framework; such mechanism keeps the interactions among agents manageable while reducing the noisy effect of the exploratory behaviors from the other agents. Between MF-$Q$ and MF-AC,  MF-$Q$ converges faster. Both FMQ and Rec-FMQ fail to reach pleasant performance, it might be because agents are essentially unable to distinguish the rewards received for the same actions, and are thus unable to update their own $Q$-values \emph{w.r.t.} the actual contributions.
It is worth noting that MAAC  is surprisingly inefficient in learning when the number of agents becomes large; it simply fails to handle the non-aggregated noises due to  the agents' explorations.



\subsection{Model-free MARL for Ising Model}

\textbf{Environment.}
In statistical mechanics, the Ising model is a mathematical framework to describe ferromagnetism \cite{ising1925beitrag}. It also has wide applications in sociophysics \cite{RePEc:eee:phsmap:v:389:y:2010:i:3:p:481-489}.
With the energy function explicitly defined, mean field  approximation \cite{stanley1971phase} is a typical way to solve the Ising model  for every spin $j$, \emph{i.e.} $\langle a^j \rangle= \sum_{a}{a^j}{P(a)}$. 
See  the  Appendix C.2 for more details. 

To fit into the MARL setting, we transform the Ising model into a stage game where the reward for each spin/agent is defined by
$r^{j} = h^j a^j + \frac{\lambda}{2} \sum_{k \in \mathcal{N}(j)} a^j a^k$; here $\mathcal{N}(j)$ is the set of nearest neighbors of spin $j$, $h^j \in \mathbb{R}$ is the external field affecting the spin $j$, and $\lambda \in \mathbb{R}$ is an interaction coefficient that determines how much the spins are motivated to stay aligned. 
Unlike the typical setting in physics, here each spin does not know the energy function, but aims to understand the environment, and to maximize its reward by learning the optimal policy  of choosing the spin state: up or down.  

In addition to the reward, the \emph{order parameter} (OP) \cite{stanley1971phase} is  a traditional measure of purity for the Ising model. OP is defined as
$\xi=\frac{|N_{\uparrow}-N_{\downarrow}|}{N}$,
where $N_{\uparrow}$ represents the number of up spins, and $N_{\downarrow}$ for the down spins. The closer the OP is to $1$, the more orderly the system is.

\begin{figure}[t]
\vskip -0.1in
\begin{subfigure}[b]{.99\linewidth}
\centering
\includegraphics[width=\columnwidth]{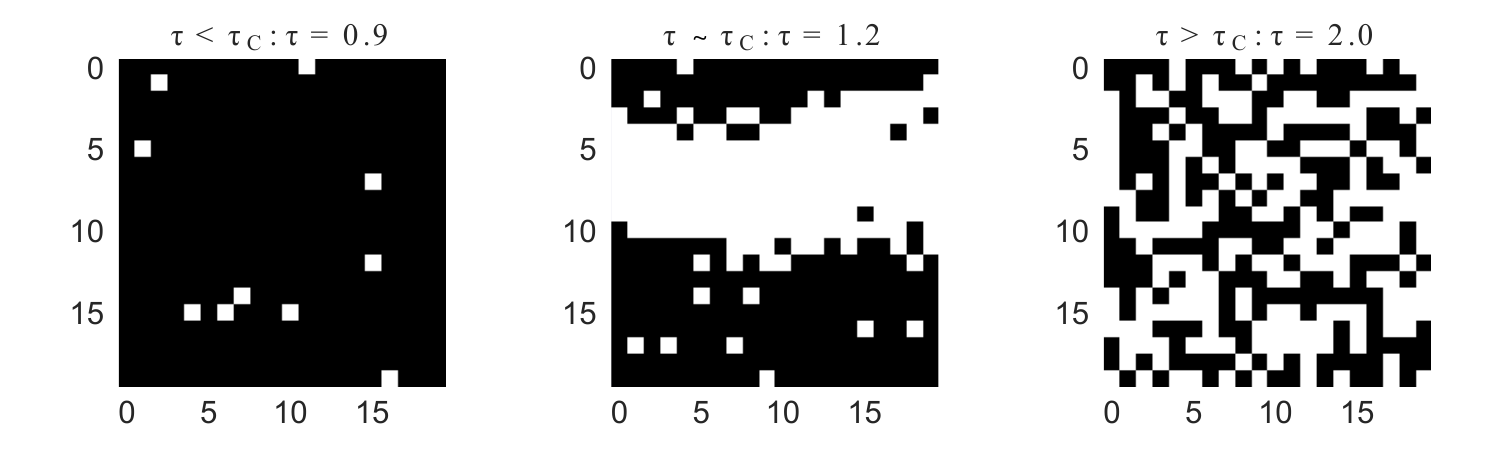}
\vskip -0.1in
\caption{MF-$Q$}
\label{subfig:mfq_pattern}
\end{subfigure}
\begin{subfigure}[b]{.99\linewidth}
\centering
\includegraphics[width=\columnwidth]{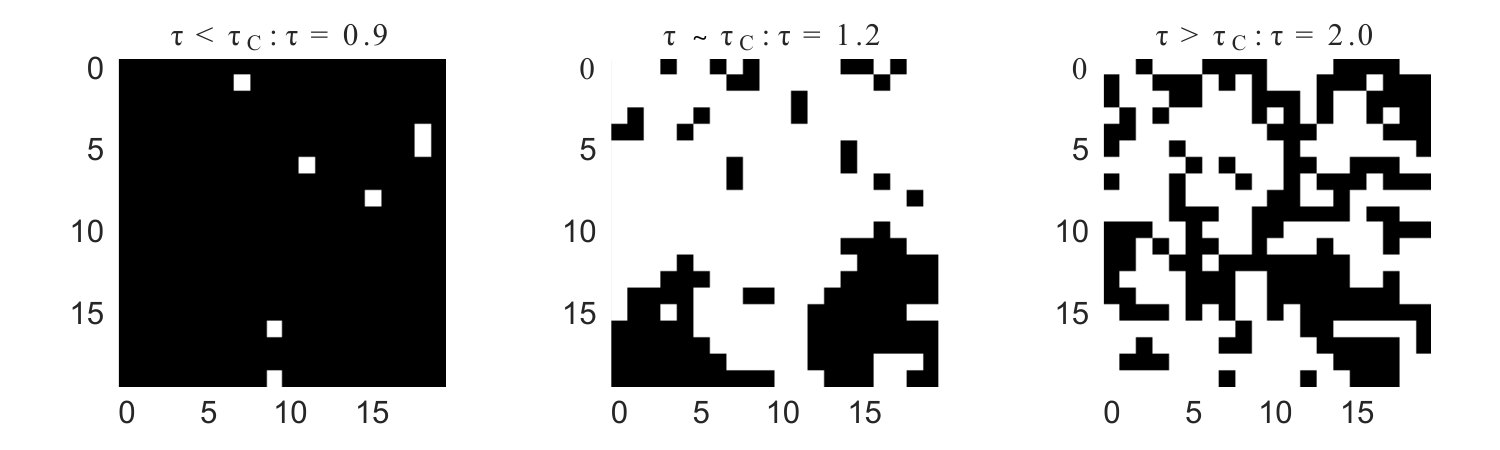}
\vskip -0.1in
\caption{MCMC}
\label{subfig:mcmc_pattern}
\end{subfigure}
\vskip -0.1in
\caption{The spins of the Ising model at equilibrium under different  temperatures.}\label{fig:ising_pattern}
\vskip -0.05in
\end{figure}


\textbf{Model Settings.}
To validate the correctness of the MF-$Q$ learning, we implement MCMC methods \cite{binder1993monte} to simulate the same Ising model and provide the ground truth for comparison. 
The full settings of MCMC and MF-$Q$ for Ising model are provided in the Appendix C.2. 
One of the learning goals is to obtain the accurate approximation of $\langle a^j \rangle$. Notice that agents here do not know exactly the energy function, 
but rather use the temporal difference learning to approximate $\langle a^j \rangle$ during the learning procedure.  Once this is accurately approximated, the Ising model as a whole should be able to converge to the  same simulation result suggested by MCMC.






\textbf{Correctness of MF-$Q$.}
Figure.~\ref{fig:OP_T} illustrates the relationship between the order parameter at equilibrium under different system temperatures.  MF-$Q$  converges nearly to the exact same plot as  MCMC, this justifies the correctness of our algorithms.  Critically, MF-$Q$ finds a similar Curie temperature (the phase change point) as MCMC that is $\tau = 1.2$. As far as we know, this is the first work that manages to solve the Ising model via model-free reinforcement learning methods.
Figure.~\ref{fig:mfq_mse} illustrates the mean squared error between the learned $Q$-value and the reward target. MF-$Q$ is shown in Fig.~\ref{subfig:mfq_400_0_8_mse} to be able to learn the target well under low temperature settings. When it comes to the Curie temperature, the environment enters into the phase change when the stochasticity dominates, resulting in a lower OP and higher MSE observed in Fig.~\ref{subfig:mfq_400_1_2_mse}.
 We visualize the equilibrium in Fig.~\ref{fig:ising_pattern}. The equilibrium points from MF-$Q$ in fact match MCMC's results under three types of temperatures. The spins tend to stay aligned under a low temperature ($\tau=0.9$). As the temperature rises ($\tau=1.2$), some spins become volatile and patches start to form as spontaneous magnetization. This phenomenon is mostly observed around the Curie temperature. After passing the Curie temperature, the system becomes  unstable and disordered due to the large thermal fluctuations, resulting in random spinning patterns.

 \begin{figure}[t]
	\vskip -0.1in
	\begin{subfigure}[b]{.42\columnwidth}
		\centering
		\includegraphics[width=.75\columnwidth]{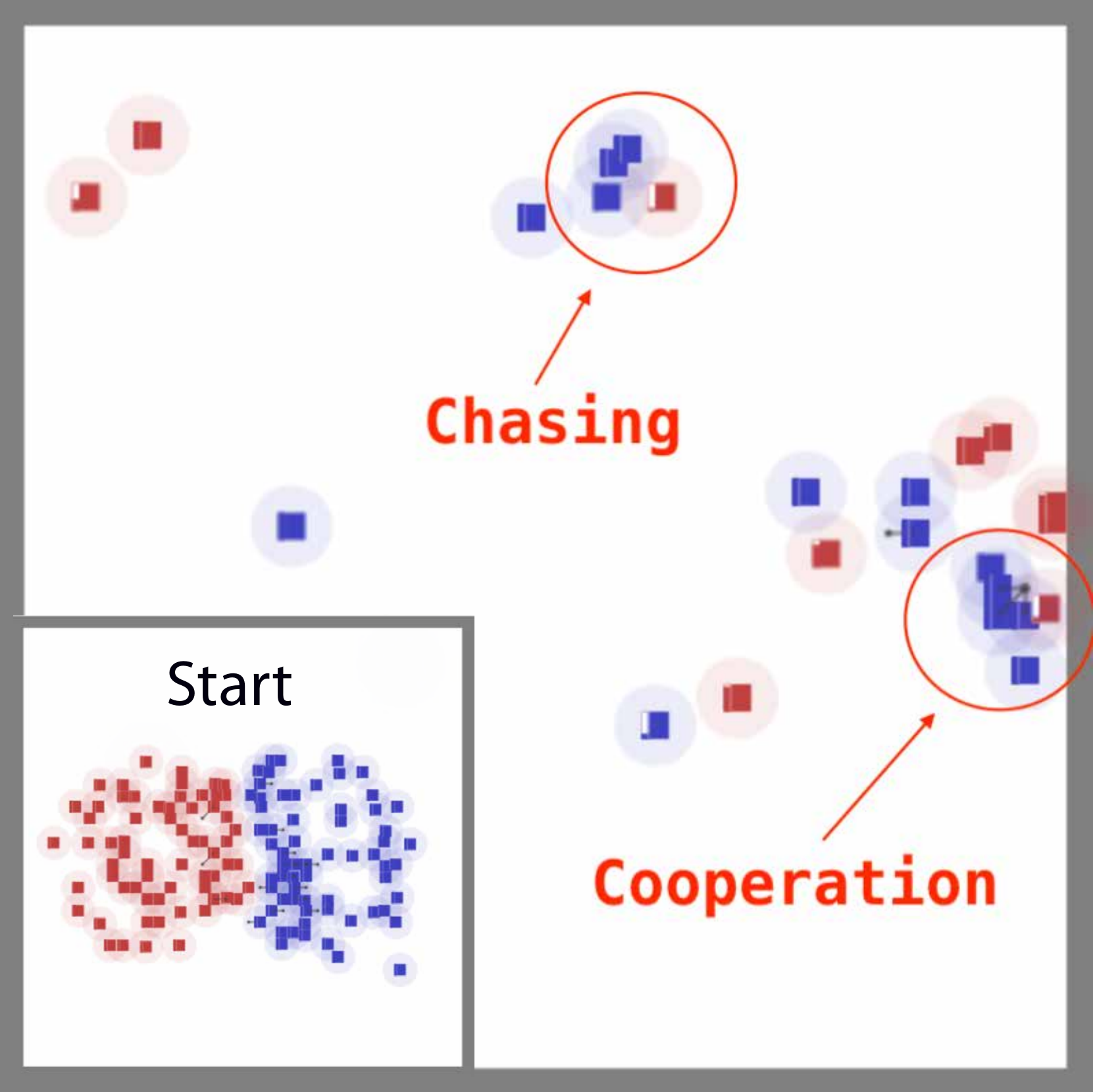}
		\vskip 0.0in
		\caption{Battle game scene.}
		\label{subfig:battle_chasing}
	\end{subfigure}
	\begin{subfigure}[b]{.56\columnwidth}
		\centering
		\includegraphics[width=0.99\columnwidth]{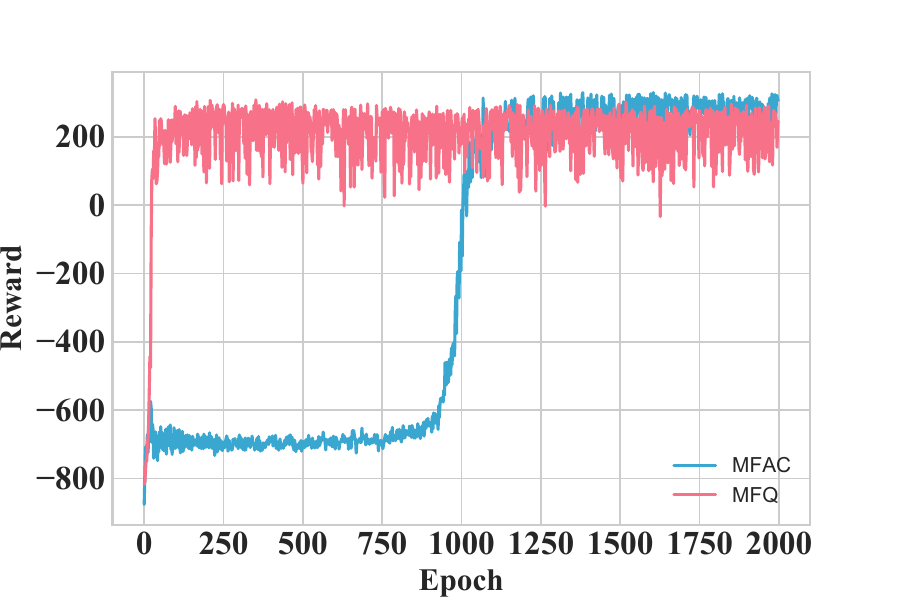}
		\vskip -0.05in
		\caption{Learning curve.}
		\label{subfig:curve_mfac_mfq}
	\end{subfigure}
	\vskip -0.1in
	\caption{The battle game: $64$ \emph{v.s.} $64$.}\label{fig:battle_es}
	\vskip -0.05in
	\end{figure}

\subsection{Mixed Cooperative-Competitive Battle Game}

\textbf{Environment.}
The Battle game in the Open-source MAgent system \cite{DBLP:conf/aaai/ZhengYCZZWY18} is a \emph{Mixed Cooperative-Competitive} scenario with two armies fighting against each other in a grid world, each empowered by a different RL algorithm.  In the setting of Fig.~\ref{subfig:battle_chasing}, each army consists of $64$ homogeneous agents. The goal of each army is to get more rewards by collaborating with teammates to destroy all the opponents. 
Agent can takes actions to either move to or attack nearby grids. Ideally, the agents army should learn skills such as chasing to hunt after training. We adopt the default reward setting: $-0.005$ for every move, $0.2$ for attacking an enemy, $5$ for killing an enemy, $-0.1$ for attacking an empty grid, and $-0.1$ for being attacked or killed.

\textbf{Model Settings.}
Our MF-$Q$ and MF-AC are compared against the baselines that are proved successful on the MAgent platform. We  focus on the battles between mean field methods (MF-$Q$, MF-AC) and their non-mean field counterparts, independent $Q$-learning (IL) and advantageous actor critic (AC). 
We exclude MADDPG/MAAC as baselines, as the framework of centralized critic cannot deal with the varying number of agents for the battle (simply because agents could  die in the battle).
Also, as we demonstrated in the previous experiment  of Fig.~\ref{fig:gsd},  MAAC  tends to scale poorly and fail when the agent number is in hundreds. 

\begin{figure}[t]
	\vskip -0.13in	
	\begin{subfigure}[b]{.49\linewidth}
		\centering
		\includegraphics[width=\columnwidth]{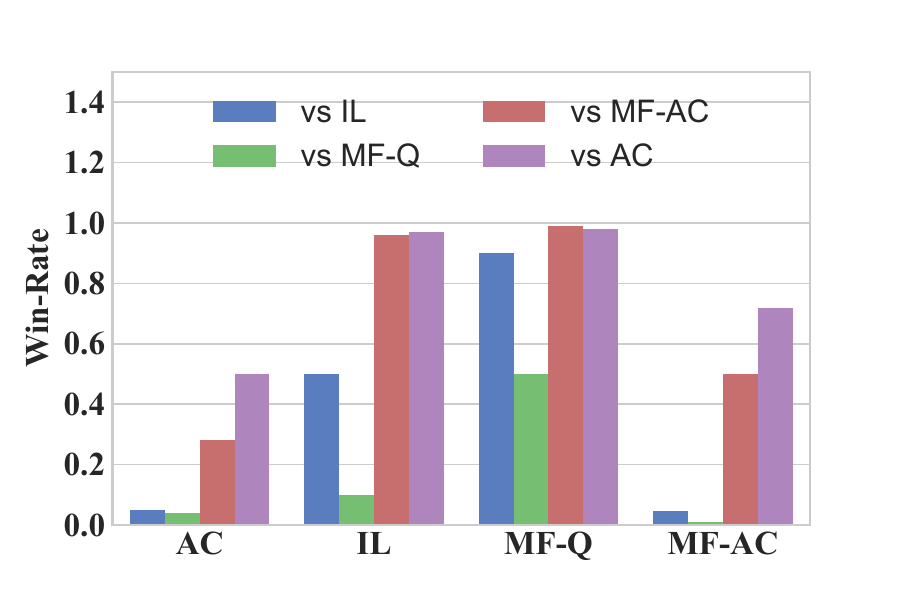}
		\vskip -0.1in
		\caption{Average wining rate.}
		\label{subfig:battle_win}
	\end{subfigure}
	\begin{subfigure}[b]{.49\linewidth}
		\centering
		\includegraphics[width=\columnwidth]{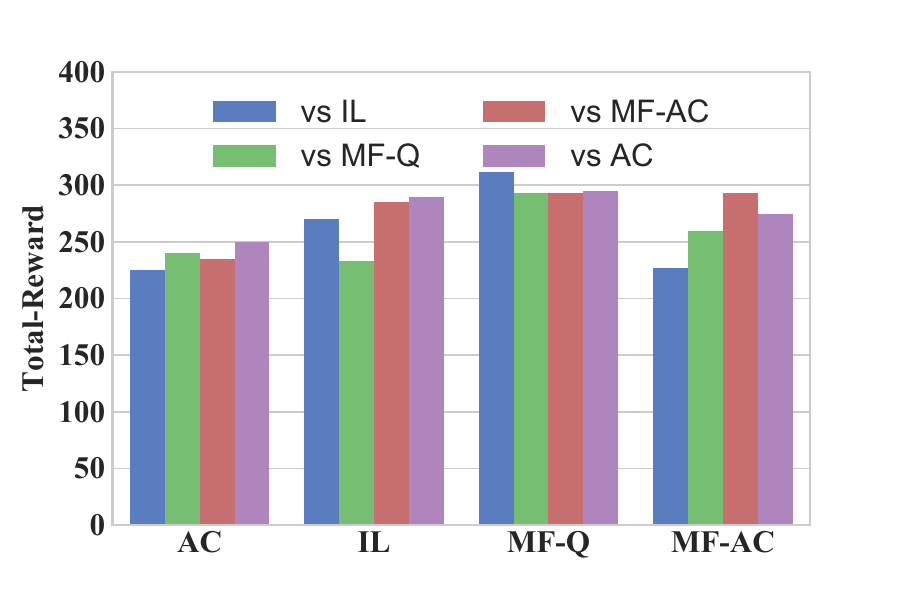}
		\vskip -0.1in
		\caption{Average total reward.}
		\label{subfig:battle_reward}
	\end{subfigure}
	\vskip -0.05in
	\caption{Performance comparisons in the battle game.}\label{fig:battle}
	\vskip -0.05in
\end{figure}

\textbf{Results and Discussion.}
We train all four models by 2000 rounds \textit{self-plays}, and then use them for comparative battles. During the training, agents can quickly pick up the skills of chasing and cooperation to kill in Fig.~\ref{subfig:battle_chasing}. 
The Fig.~\ref{fig:battle} shows the result of winning rate and the total reward over 2000 rounds cross-comparative experiments. It is evident that on all the metrics mean field methods, MF-$Q$ largely outperforms the corresponding baselines, \emph{i.e.} IL and AC respectively, which shows the effectiveness of the mean field MARL algorithms. Interestingly, IL performs far better than AC and MF-AC (2nd block from the left in Fig.~\ref{subfig:battle_win}), although it is worse than the mean field counterpart MF-$Q$. This might imply the effectiveness of off-policy learning with shuffled buffer replay in many-agent RL towards a more stable learning process. Also, the $Q$-learning family tends to introduce a positive bias \cite{hasselt2010double} by using the maximum action value as an approximation for the maximum expected action value, and such overestimation can be beneficial for each single agent to find the best response to others even though the environment itself is still changing. On the other hand, On-policy methods  need to comply with the GLIE assumption (Assumption 2 in Sec 3.3) so as to converge properly to the optimal value \cite{singh2000convergence}, which is in the end a greedy policy as off-policy methods.
Figure.~\ref{subfig:curve_mfac_mfq} further shows the self-play learning curves  of MF-AC and MF-$Q$. MF-$Q$ presents a faster convergence speed than MF-AC, which is consistent with the findings in the Gaussian Squeeze task (see Fig.~\ref{subfig:gsd_500} \& \ref{subfig:gsd_1000}).
Apart from 64, we further test the scenarios when the agent size is 8, 144, 256, the comparative results keep the same relativity as Fig.~\ref{fig:battle}; we omit the presentations for clarity.




\section{Conclusions}
In this paper, we developed mean field reinforcement learning methods to  model the dynamics of interactions in the multi-agent systems. 
MF-$Q$ iteratively learns each agent's best response to the mean effect from its neighbors; this effectively transform the many-body problem into a two-body problem. 
Theoretical analysis on the convergence of the MF-$Q$  algorithm to Nash $Q$-value  was provided. 
Three types of tasks have justified the effectiveness of our approaches. In particular, we report the first result to solve the Ising model using model-free reinforcement learning methods.



\section*{Acknowledgement}

We sincerely thank Ms. Yi Qu for her generous help on the graphic design.

\bibliographystyle{icml2018}
\bibliography{mfrl}

\newpage
\onecolumn
\appendix 

\section{Detailed mean field reinforcement learning algorithms}

We published the code at \url{https://github.com/mlii/mfrl}.
\label{appendix:mfrl_algo}

{\small
\vskip 0.1in
\begin{algorithm*}
	\caption{Mean Field $Q$-learning (MF-$Q$)} \label{algo:mf-q}
	\begin{algorithmic}
		\STATE Initialise $Q_{\phi^j}$, $Q_{\phi^j_{-}}$, and $\bar{a}^j$ for all $j \in \{1,\dots,N\}$
		\WHILE{training not finished}
		\FOR{$m=1,..., M$}
		\STATE For each agent $j$, sample action $a^j$ from $Q_{\phi^j}$ by Eq.~\eqref{boltzman_q}, with the current mean action $\bar{a}^j$ and the exploration rate $\beta$
		\STATE For each agent $j$, compute the new mean action $\bar{a}^j$ by Eq.~\eqref{forward_eq}
		\ENDFOR
		\STATE Take the joint action $\vs{a} = [a^1,\dots,a^N]$ and observe the reward $\vs{r} = [r^1,\dots,r^N]$ and the next state $s'$
		\STATE Store $\langle s,\vs{a},\vs{r},s',\bar{\vs{a}} \rangle$ in replay buffer $\mathcal{D}$, where $\bar{\vs{a}} = [\bar{a}^1,\dots,\bar{a}^N]$
		\FOR{$j=1\mbox{~to~}N$}
		\STATE Sample a minibatch of $K$ experiences $\langle s,\vs{a},\vs{r},s',\bar{\vs{a}} \rangle$ from $\mathcal{D}$
		\STATE Sample action $a^j_{-}$ from $Q_{\phi^j_{-}}$ with $\bar{a}^j_{-} \gets \bar{a}^j$
		\vskip -0.05in
		\STATE Set $y^j = r^j+\gamma\, v^{\mf}_{\phi^j_{-}}(s')$ by Eq.~\eqref{mfv}
		\vskip -0.10in
		\STATE Update the $Q$-network by minimizing the loss $\mathcal{L}(\phi^j)=\frac{1}{K}\sum\big(y^j-Q_{\phi^j}(s^j, a^j,\bar{a}^j)\big)^2$
		\ENDFOR     
		\STATE Update the parameters of the target network for each agent $j$ with learning rate $\tau$:
		$$
		\phi^j_-\gets\tau\phi^j+(1-\tau)\phi^j_-
		$$
		\ENDWHILE
	\end{algorithmic}
\end{algorithm*}
}%

	\vskip 0.2in
\begin{algorithm*}
	\caption{Mean Field Actor-Critic (MF-AC)}\label{algo:mf-ac}
 \begin{algorithmic}
	\STATE Initialize $Q_{\phi^j}$, $Q_{\phi^j_{-}}$, $\pi_{\theta^j}$, $\pi_{\theta^j_{-}}$, and $\bar{a}^j$ for all $j \in \{1,\dots,N\}$
	\WHILE{training not finished}
	\STATE For each agent $j$, sample action $a^j=\pi_{\theta^j}(s)$; compute the new mean action $\bar{\vs{a}} = [\bar{a}^1,\dots,\bar{a}^N]$
	\STATE Take the joint action $\vs{a} = [a^1,\dots,a^N]$ and observe the reward $\vs{r} = [r^1,\dots,r^N]$ and the next state $s'$
	\STATE Store $\langle s,\vs{a},\vs{r},s',\bar{\vs{a}} \rangle$ in replay buffer $\mathcal{D}$
	\FOR{$j=1\mbox{~to~}N$}
	\STATE Sample a minibatch of $K$ experiences $\langle s,\vs{a},\vs{r},s',\bar{\vs{a}} \rangle$ from $\mathcal{D}$
	\vskip .05in
\STATE Set $y^j = r^j+\gamma\, v^{\mf}_{\phi^j_{-}}(s')$ by Eq.~\eqref{mfv}
	\STATE Update the critic by minimizing the loss $\mathcal{L}(\phi^j)=\frac{1}{K}\sum\big(y^j-Q_{\phi^j}(s, a^j,\bar{a}^j)\big)^2$
	\STATE Update the actor using the sampled policy gradient:
	\vskip -0.1in
	$$
	\nabla_{\theta^j}\mathcal{J}(\theta^j)\approx\frac{1}{K}\sum\nabla_{\theta^j}\log\pi_{\theta^j}(s')Q_{\phi^j_{-}}(s', a^j_{-}, \bar{a}^j_{-})\Big|_{a^j_{-}=\pi_{\theta^j_{-}}(s')}$$
	\ENDFOR     
	\STATE Update the parameters of the target networks for each agent $j$ with learning rates $\tau_{\phi}$ and $\tau_{\theta}$:
	\begin{align*}
		\phi^j_- &\gets \tau_{\phi}\phi^j+(1-\tau_{\phi})\phi^j_-\\
		\theta^j_- &\gets \tau_{\theta}\theta^j+(1-\tau_{\theta})\theta^j_-
	\end{align*}
	\ENDWHILE
\end{algorithmic}
\end{algorithm*}

\clearpage
\section{Proof of the bound for the remainder term in Eq. \ref{mean-field-first}} \label{2ndbound}

Recall Eq. \eqref{mean-field-final} that we approximate the action $a^k$ taken by the neighboring agent $k$ with the mean action $\bar{a}$ calculated from the neighborhood $\mathcal{N}(j)$. The state $s$ and the action $a^j$ of the central agent $j$ can be considered as fixed parameters; the indices $j,k$ of agents are essentially irrelevant to the derivation. With those omitted for simplicity, We rewrite the expression of the pairwise $Q$-function as $Q(a) \triangleq Q^j(s, a^j, a^k)$.

Suppose that $Q$ is $M$-smooth, where its gradient $\nabla Q$ is Lipschitz-continuous with constant $M$ such that for all $a, \bar{a}$
\begin{align}
\| \nabla Q(a) - \nabla Q(\bar{a}) \|_2 \le M \| a - \bar{a} \|_2,
\end{align}
where $\| \cdot \|_2$ indicates the $\ell_2$-norm.

With the Lagrange's mean value theorem, we have
\begin{align}
\nabla Q(a) - \nabla Q(\bar{a}) = \nabla Q(\bar{a} + 1 \cdot (a - \bar{a})) - \nabla Q(\bar{a}) = \nabla^2 Q(\bar{a} + \epsilon \cdot (a - \bar{a})) \cdot (a - \bar{a}), \mbox{~~~~where~~} \epsilon \in [0,1].
\end{align}
Take the $\ell_2$-norm on the both sides of the above equation, it follows from the smoothness condition that
\begin{align}
\| \nabla Q(a) - \nabla Q(\bar{a}) \|_2 = \| \nabla^2 Q(\bar{a} + \tau \cdot (a-\bar{a})) \cdot (a - \bar{a}) \|_2 \le M \| a - \bar{a} \|_2.
\end{align}
Define $\delta{a} \triangleq a - \bar{a}$ and the normalized vector $\delta{\hat{a}} \triangleq \nicefrac{a - \bar{a}}{\| a - \bar{a} \|_2}$ with $\| \delta{\hat{a}} \|_2 = 1$, it follows from the above inequality 
\begin{align}
\| \nabla^2 Q(a + \tau \cdot \delta{a}) \cdot \delta{\hat{a}} \|_2 \le M.
\end{align}
By arbitrary choice of (the unnormalized vector) $\delta{a}$ such that the magnitude $\|\delta{a}\|_2 \to 0$, it follows from above that
\begin{align}
\| \nabla^2 Q(a) \cdot \delta{\hat{a}} \|_2 \le M.
\end{align}
By aligning (the normalized vector) $\delta{\hat{a}}$ in the direction of the eigenvectors of the Hessian matrix $\nabla^2 Q$, we can obtain for any eigenvalue $\lambda$ of $\nabla^2 Q$ that
\begin{align}
\| \nabla^2 Q(a) \cdot \delta{\hat{a}} \|_2 = \| \lambda \cdot \delta{\hat{a}} \|_2 = |\lambda| \cdot \| \delta{\hat{a}} \|_2 \le M,
\end{align}
which indicates that all eigenvalues of $\nabla^2 Q$ can be bounded in the symmetric interval $[-M, M]$.

As the Hessian matrix $\nabla^2 Q$ is real symmetric and hence diagonalizable, there exist an orthogonal matrix $U$ such that $U^\top [\nabla^2 Q] U = \Lambda \triangleq \diag[\lambda_1,\dots,\lambda_D]$. It then follows that
\begin{align}
\delta{a} \cdot \nabla^2 Q \cdot \delta{a} = [U\delta{a}]^\top \Lambda [U\delta{a}] = \sum_{i=1}^D \lambda_i [U\delta{a}]_i^2, \mbox{~~~~with~~} -M \| U\delta{a} \|_2 \le \sum_{i=1}^D \lambda_i [U\delta{a}]_i^2 \le M \| U\delta{a} \|_2
\end{align}
Recall the definition $\delta{a} = a - \bar{a}$ in Eq. \eqref{eq:abar}, where $a$ is the one-hot encoding for $D$ actions, and $\bar{a}$ is a $D$-dimensional multinomial distribution. It can be shown that
\begin{align}
\| U\delta{a} \|_2 = \| \delta{a} \|_2 = (a - \bar{a})^\top (a - \bar{a}) = a^\top a + \bar{a}^\top \bar{a} - \bar{a}^\top a - a^\top \bar{a} = 2(1- \bar{a}_i) \le 2,
\end{align}
where $i$ represents the specific action $a$ has represented such that $a_{i'} = 0$ for $i' \neq i$.

With all elements assembled, we have proved that each single remainder term $R^j_{s, a^j}(a^k)$ in Eq. \eqref{mean-field-final} is bounded in $[-2M, 2M]$.

\clearpage
\section{Experiment details}

 \subsection{Gaussian Squeeze}
 \label{appendix:GS}


\textbf{IL, FMQ, Rec-FMQ and MF-$Q$} all use a three-layer MLP to approximate $Q$-value. All agents share the same $Q$-network for each experiment. The shared $Q$-network takes an agent embedding as input and computes $Q$-value for each candidate action. For MF-$Q$, we also feed in the action approximation $\bar{a}$. We use the Adam optimizer with a learning rate of 0.00001 and $\epsilon$-greedy exploration unless otherwise specified. For FMQ, we set the exponential decay rate $s=0.006$, start temperature \emph{max\_temp=1000} and FMQ heuristic $c=5$. For Rec-FMQ, we set the frequency learning rate $\alpha_f=0.01$.

\textbf{MAAC and MF-AC} use the Adam optimizer with a learning rate of 0.001 and 0.0001 for Critics and Actors respectively, and $\tau=0.01$ for updating the target networks. We share the Critic among all agents in each experiment and feed in an agent embedding as extra input. Actors are kept separate. The discounted factor $\gamma$ is set to be 0.95 and the mini-batch size is set to be 200. The size of the replay buffer is $10^6$ and we update the network parameters after every 500 samples added to the replay buffer.

For all models, we use the performance of the joint-policy learned up to that point if learning and exploration were turned off (\emph{i.e.}, take the greedy action w.r.t. the learned policy) to compare our method with the above baseline models.

\subsection{Ising Model}
\label{appendix:ising}

An Ising model is defined as a stateless system with $N$ homogeneous sites on a finite square lattice. Each site determines their individual spin $a^j$ to interact with each other and aims to minimize the system energy for a more stable environment. The system energy is defined as 
\begin{align}
\label{appendix:ising-energy}
    E(a, h) = -\sum_j{(h^j a^j + \frac{\lambda}{2} \sum_{k \in \mathcal{N}(j)} a^j a^k)}
\end{align}
where $\mathcal{N}(j)$ is the set of nearest neighbors of site $j$, $h^j \in \mathbb{R}$ is the external field affecting site $j$, and $\lambda \in \mathbb{R}$ is an interaction term determines how much the sites tend to align in the same direction. The system is said to reach an equilibrium point when the system energy is minimized, with the probability 
\begin{align}
\label{appendix:ising-equilibrium}
    & P(a) = \frac{\exp\left({-E(a, h)}/{\tau}\right)}{\sum_a {\exp({-E(a, h)}/{\tau}) } },
\end{align}
where $\tau$ is the system temperature. When the temperature rises beyond a certain point (the Curie temperature), the system can no longer keep a stable form and a phase transition happens. As the ground-truth is known, we would be able to evaluate the correctness of the $Q$-function learning when there is a large body of agents interacted.

The mean field theory provides an approximate solution to $\langle a^j \rangle= \sum_{a}{a^j}{P(a)}$ through a set of self-consistent mean field equations
\begin{align}
\label{appendix:mean-field-equations}
    & \langle a^j \rangle = \frac{\exp\left(-[h^j a^j + \lambda \sum_{k \in \mathcal{N}(j)} \langle a^k \rangle]/{\tau}\right)}{1+\exp\left(-[h^j a^j + \lambda \sum_{k \in \mathcal{N}(j)} \langle a^k \rangle]/{\tau}\right)}.
\end{align}
which can be solved iteratively by
\begin{align}
\label{appendix:mean-field-iterative-solution}
    & \langle a^j \rangle^{(t+1)} = \frac{\exp\left(-[h^j a^j + \lambda \sum_{k \in \mathcal{N}(j)} \langle a^k \rangle^{(t)}]/{\tau}\right)}{1+\exp\left(-[h^j a^j + \lambda \sum_{k \in \mathcal{N}(j)} \langle a^k \rangle^{(t)}]/{\tau}\right)},
\end{align}
where $t$ represents the number of iterations.

\begin{algorithm*}[t]
\caption{MCMC in Ising Model}
\begin{algorithmic}
\STATE initialize spin state $\vs{a} \in \{-1,1\}^N$ for $N$ sites
\WHILE{training not finished}
\STATE randomly choose site $j \in \mathcal{N}(j)$
\STATE flip the spin state for site $j$: $a^j_{-} \gets -a^j$
\STATE compute neighbor energy 
$E(a, h) = -\sum_j{(h^j a^j + \frac{\lambda}{2} \sum_{k \in \mathcal{N}(j)} a^j a^k)}$ for $a^j$ and $a^j_{-}$
\STATE randomly choose $\epsilon \sim U(0, 1)$
\IF{$\exp({{(E(a^j,h)-E(a^j_{-},h))}/{\tau}}) > \epsilon$}
\STATE $a^j \gets a^j_{-}$
\ENDIF
\ENDWHILE
\end{algorithmic}
\end{algorithm*}

To learn an optimal joint policy $\vs{\pi}^*$ for Ising model, we use the stateless $Q$-learning with mean field approximation (MF-$Q$), defined as
\begin{align}
\label{stateless-mfq}
      & Q^j({a^j},\bar{a}^j)\gets Q^j({a^j},\bar{a}^j)+\alpha[r^j-Q^j({a^j},\bar{a}^j)],
\end{align}
where the mean $\bar{a}^j$ is given as the mean $\langle a^j \rangle$ from the last time step, and the individual reward is
\begin{align}
      & r^{j} = h^j a^j + \frac{\lambda}{2} \sum_{k \in \mathcal{N}(j)} a^j a^k.
\end{align}
To balance the trade-off between exploration and exploitation under low temperature settings, we use a policy with Boltzmann exploration and a decayed exploring temperature. The temperature for Boltzmann exploration of MF-$Q$ is multiplied by a decay factor exponentially through out the training process.

Without lost of generality, we assume $\lambda>0$, thus neighboring sites with the same action result in lower energy (observe higher reward) and are more stable. Each site should also align with the sign of external field $h^j$ to reduce the system energy. For simplification, we eliminate the effect of external fields and assume the model to be discrete, \emph{i.e.}, $\forall j \in N, h^j=0, a^j \in \{-1,1\}$.

We simulate the Ising model using Metropolis Monte Carlo methods (MCMC). After initialization, we randomly change a site's spin state and calculate the energy change, select a random number between 0 and 1, and accept the state change only if the number is less than $e^{\frac{(E^j-E^j_{-})}{\tau}}$. This is called the Metropolis technique, which saves computation time by selecting the more probable spin states.


\subsection{Battle Game}


\textbf{IL and MF-$Q$} have almost the same hyper-parameters settings. The learning rate is $\alpha=10^{-4}$, and with a dynamic exploration rate linearly decays from $\gamma=1.0$ to $\gamma=0.05$ during the 2000 rounds training. The discounted factor $\gamma$ is set to be 0.95 and the mini-batch size is 128. The size of replay buffer is $5 \times 10^5$.

\textbf{AC and MF-AC} also have almost the same hyper-parameters settings. The learning rate is $\alpha=10^{-4}$, the temperature of soft-max layer in $actor$ is $\tau=0.1$. And the coefficient of entropy in the total loss is 0.08, the coefficient of value in the total loss is 0.1.

\clearpage
\section{Further details towards the theoretical guarantee of MF-$Q$}\label{mfq_detail}

\bolzproposition*

\begin{proof}
Following the contraction mapping theorem \cite{kreyszig1978introductory}, in order to be a contraction, the operator has to satisfy:
\[
d(\mathcal{B}(\vs{a}), \mathcal{B}(\vs{b})) \leq \alpha d(\vs{a}, \vs{b}),~~~\forall \vs{a}, \vs{b}
\]
where $0\leq \alpha < 1$ and $\mathcal{B}(\vs{a}) \triangleq [\mathcal{B}(a^1),\dots,\mathcal{B}(a^N)]$.

Here we start from binomial case and then adapt to the multinomial case in general.
We first rewrite $\mathcal{B}(a^j)$ as
\begin{align}
\mathcal{B}(a^j) = \pi^j(a^j | s, \bar{a}^j) =& \frac{\exp{\big(- \beta Q^j_{t}(s, a^j, \bar{a}^j)\big)}}{\exp{\big(- \beta Q^j_{t}(s, a^j, \bar{a}^j)\big) + \exp{\big(- \beta Q^j_{t}(s, \neg a^j, \bar{a}^j)\big)}}} \notag \\
=& \frac{1}{1+\exp{(- \beta \cdot \Delta Q(s, a^j, \bar{a} ))}}~,
\end{align}
where  $\Delta Q(s, a^j, \bar{a} ) =  Q(s, {a^{\neg}}^j, \bar{a} ) - Q(s, a^j, \bar{a} )$. 

Then we have
\begin{align}
|\mathcal{B}(a^j) - \mathcal{B}(b^j)| =& \left| \frac{1}{1+e^{{- \beta \cdot \Delta Q(s, a^j, \bar{a})}}} - \frac{1}{1+e^{{- \beta \cdot \Delta Q(s, b^j, \bar{a})}}}\right| \nonumber \\
=&\left| \dfrac{\beta e^{- \beta \Delta Q_0}}{(1+e^{- \beta \Delta Q_0})^2}     \right| \left| \Delta Q(s, a^j, \bar{a}) - \Delta Q(s, b^j, \bar{a})  \right| \nonumber \\
 \leq & \frac{1}{4T}\cdot \left| Q(s, {a^{\neg}}^j, \bar{a} )  - Q(s, {b^{\neg}}^j, \bar{a} ) + Q(s, b^j, \bar{a} ) - Q(s, a^j, \bar{a} ) \right| \nonumber \\
 \leq & \frac{1}{4T}\cdot \left( K \cdot \left| 1- a^j - (1 - b^j) \right| + K \cdot \left| a^j - b^j \right|       \right) \nonumber \\
\leq & \frac{1}{4T}\cdot 2K \cdot \sum_{j} \left| a^j - b^j \right|~.
\end{align}
In the second equation, we apply the mean value theorem in calculus:  $\exists x_0 \in [x_1, x_2]$, \emph{s.t.},
$
f(x_1) - f(x_2) = f'(x_0)(x_1 - x_2)
$. In the third equation we use the maximum value for $e^{- \beta \Delta Q_0}/({1+e^{- \beta \Delta Q_0}})^2 = 1/4$ when $Q_0=0$. In the last equation we apply the Lipschitz constraint in the assumption where constant $K \geq 0$. Finally, we have:
\begin{align}
d(\mathcal{B}(\vs{a}), \mathcal{B}(\vs{b})) &\leq \frac{1}{4T}\cdot 2K \cdot \sum_{j} \left| a^j - b^j \right|  \nonumber \\
&= \frac{K}{2T}d(\vs{a}, \vs{b})
\end{align}
In order for the contraction to hold, $T>\frac{K}{2}$. In other words, when the action space is binary for each agent, and the temperature is sufficiently large, the mean field procedure converges.

This proposition can be easily extended to multinomial case by replacing binary variable $a^j$ by a multi-dimensional binary indicator vector $\vs{a}^j$, on each dimension,  the rest of the derivations would remain essentially the same.
\end{proof}

\subsection{Discussion on Rationality}

In line with \cite{bowling2001rational, bowling2002multiagent}, we argue that to better evaluate a multi-agent learning algorithm, on top of the convergence guarantee, discussion on property of Rationality is also needed.

\begin{property} (also see \cite{bowling2001rational, bowling2002multiagent})
In an $N$-agent stochastic game defined in this paper, given all agents converge to stationary policies, if the learning algorithm converges to a policy that is a \textbf{best response} to the other agents' policies, then the algorithm is Rationale.
\end{property}

Our mean field $Q$-learning is rational in that Eq.~(\ref{pairwise-Q}) converts many agents interactions into two-body interactions between a single agent and the distribution of other agents actions. When all agents follow stationary policies, their policy distribution would be stationary too. As such the two-body stochastic game becomes an MDP, and the agent would choose a policy (based on Assumption \ref{glieassum}) which is the best response to the distribution of other stationary policies. As agents are symmetric in our case, they all show the best response to the distributions, and are therefore rational.

\section{Proof of Mean Field Reinforcement Learning with Function Approximation}

Previous convergence results in Theorem.1 has shown that the Mean Field Q-learning algorithm will converge when the Q function is in a tabular case. We now move onto the proof that the MF-Q algorithm will converge when the Q function is represented by some functional approximations.

An $N$-agent (or, $N$-player) stochastic game $\Gamma$ is formalized by the tuple $\Gamma \triangleq \big( \mathcal{S}, \mathcal{A}^1, \dots, \mathcal{A}^N, r^1, \dots, r^N, p, \gamma \big)$. The state-space $\mathcal{S}$ is finite.
Let $(\mathcal{S}, p_\vs{\pi})$ be the Markov chain induced by the joint policy $\vs{\pi}$, and we assume it to be uniformly ergodic.

Let $\mathcal{Q}=\{\vs{Q}_\theta \}$ be a family of real-valued functions defined on $\mathcal{S}\times \mathcal{A} \times \bar{\mathcal{A}}$, where $\bar{\mathcal{A}}$ is the action space for the mean actions computed from the neighbors.
Assuming that the function class is linearly parameterized, for each agent $j$, Q can be expressed as the linear span of a fixed set of $P$ linearly independent functions $\omega_p^j: \mathcal{S}\times \mathcal{A} \times \bar{\mathcal{A}} \rightarrow \mathbb{R}$. Given the parameter vector $\phi^j \in \mathbb{R}^P$, for each agent, the function $Q_{\phi^j}$ is thus defined as 
\begin{equation}
Q_{\phi^j}(s, a^j, \bar{a}^j)=\sum_{p=1}^P \omega_p^j(s, a^j, \bar{a}^j) \phi^j(p) = \omega^j(s, a^j, \bar{a}^j)^\top \phi^j
\end{equation}

In the functional approximation setting, we can apply the update rules:
\begin{align}
\phi^j_{t+1}&= \phi^j_t + \alpha_t \Delta_t   \nabla_{\phi^j}{Q_{\phi^j}(s, a^j, \bar{a}^j)} \nonumber \\
&= \phi^j_t + \alpha_t \Delta_t \omega^j(s, a_t^j, \bar{a}_t^j) \label{itersteps}
\end{align}

In the above, $\Delta_t$ is the temporal difference at time $t$.
\begin{align}
\Delta_t &=  r^j+\gamma\, v^{\mf}_{\phi^j}(s') - {Q_{\phi^j}(s, a^j, \bar{a}^j)} \\
&= r^j+\gamma\, \mathbb{E}_{\vs{a} \sim \vs{\pi}^{\mf}} \big[ Q_{\phi^j}(s', a^j, \bar{a}^j) \big] - Q_{\phi^j}(s, a^j, \bar{a}^j).
\end{align}
And the goal is to derive the parameter vector $\vs\phi = \{\phi^j \}$ such that ${\vs\omega}^{\top} \vs\phi$  approximates the (local) Nash Q-values. At each time step, the learning policy $\vs\pi_{\phi_t}$ is the Botlzmann policy with respect to ${\vs\omega}^{\top} \vs\phi$. Give the \textbf{Proposition 1}, we know that the policy $\vs{\pi}_{\phi_t}$ is $\frac{K}{2T}$ Lipschitz continuous with respect to $\phi_t$.

Similar to the framework used in the convergence proof of Q-learning with function approximation \cite{melo2008analysis}, we establish convergence of Eq.~\eqref{itersteps} by adopting an ordinary differentiable equation (ODE) with a globally asymptotically stable equilibrium point where the trajectories closely follow. 

%


\begin{theorem}

Given the MDP $\Gamma$, $\vs{\pi}_{\phi_t}$, $\{\vs\omega_p, p =1, ..., P \}$, and the learning policy  $\vs{\pi}_{\phi_t}$ that is $\frac{K}{2T}$ Lipschitz continuous with respect to $\phi_t$, if the Assumptions \ref{tableassum}, \ref{glieassum} \& \ref{2nasheq}, and Lemma \ref{fundamental}'s first and second conditions are met, then there exists $C_0$ such that the algorithm in Eq.~\eqref{itersteps} converges w.p.1 if $\frac{K}{2T} < C_0$.
\end{theorem}

\begin{proof}
We first re-write the Eq.~\eqref{itersteps} as on ODE:
\begin{align}
\dfrac{d\vs\phi}{dt}& = \mathbb{E}_{\vs\phi}\left[\vs\omega_s^{\top}\left(\vs{r}(s, \vs\mathbf{a}, s') + \gamma \vs\omega_{s'}^{\top}\vs\phi  - \vs\omega_{s}^{\top}\vs\phi  \right)  \right]	\nonumber \\
& =  \mathbb{E}_{\vs\phi}\left[\vs\omega_s^{\top}(\gamma \vs\omega_{s'}^{\top} - \vs\omega_{s}^{\top} )\right] \vs\phi + \mathbb{E}_{\vs\phi}\left[\vs\omega_s^{\top}(\vs{r}(s, \vs\mathbf{a}, s'))\right] \nonumber \\
& = \vs{A}_{\phi} \vs\phi + \vs{b}_{\phi} 
\label{ode}
\end{align}
Notice that we use a vector for considering the updating rule for the Q function of  each agent. We can easily know that necessity condition of the equilibrium is that it must follow $\vs\phi^* = \vs{A}_{\phi^*}^{-1}\vs{b}_{\phi^*}$. The existence of the such equilibrium has been restricted in the scenario that meets Assumption $3$. In the proof of Theorem 1 we have already pointed out that under the Assumption $3$, the existing equilibrium, either in the form of global equilibrium or in the form of saddle-point equilibrium, is unique.

 Let $\tilde{\vs\phi} = \vs\phi_t - \vs\phi^*$, we have:
\begin{align}
\dfrac{d}{dt} || \tilde{\vs\phi}||_{2}  & = 2 \vs\phi \cdot \frac{d\vs\phi}{dt} - 2 \vs\phi^*\cdot \frac{d\vs\phi}{dt}  \nonumber \\
&= 2 \tilde{\vs\phi}^{\top}(\vs{A}_{\phi} \vs\phi + \vs{b}_{\phi} -\vs{A}_{\phi^*} \vs\phi^* - \vs{b}_{\phi^*}) \nonumber \\
& = 2 \tilde{\vs\phi}^{\top}(\vs{A}_{\phi^*} \vs\phi - \vs{A}_{\phi^*} \vs\phi + \vs{A}_{\phi} \vs\phi + \vs{b}_{\phi} -\vs{A}_{\phi^*} \vs\phi^* - \vs{b}_{\phi^*})  \nonumber \\
& = 2 \tilde{\vs\phi}^{\top}\vs{A}_{\phi^*} \tilde{\vs\phi} + 2 \tilde{\vs\phi}^{\top}(\vs{A}_{\phi} - \vs{A}_{\phi^*})\vs\phi + 2 \tilde{\vs\phi}^{\top}(\vs{b}_{\phi} - \vs{b}_{\phi^*}) \nonumber \\
& \leq  2 \tilde{\vs\phi}^{\top} \left(\vs{A}_{\phi^*} + \sup_{\vs\phi} || \vs{A}_{\phi} - \vs{A}_{\phi^*}||_2 + \sup_{\vs\phi} \dfrac{||\vs{b}_{\phi} - \vs{b}_\phi^*||_2}{||\vs\phi - \vs\phi^*||_2}   \right) \tilde{\vs\phi}
\label{odefinal}
\end{align}
As we know that the policy $\vs\pi_{\phi_t}$ is Lipschitz w.r.t $\phi_{t}$, this implies that $\vs{A}_{\phi}$ and $\vs{v}_{\phi}$ are also Lipschitz continuous w.r.t to $\phi$. In other words, if $\frac{K}{2T} \leq C_0$  is sufficiently small and close to zero, 
then the norm term of $\left(\sup_{\vs\phi} || \vs{A}_{\phi} - \vs{A}_{\phi^*}||_2 + \sup_{\vs\phi} \dfrac{||\vs{b}_{\phi} - \vs{b}_\phi^*||_2}{||\vs\phi - \vs\phi^*||_2}\right)$ goes to zero. Considering near the equilibrium point $\phi^*$, $\vs{A}_{\phi^*}$ is a negative definite matrix, the Eq.~\eqref{odefinal} tends to be negative definite as well, so the ODE in Eq.\eqref{ode} is globally asymptotically stable and the conclusion of the theorem follows.
\end{proof}

\end{document}